\documentclass[jgrga,draft]{agutex}
\usepackage{graphicx}
\usepackage{tabularx}
\usepackage{lineno}
\usepackage{epstopdf}
\usepackage{rotating}

\setkeys{Gin}{draft=false}
\authorrunninghead{YEO ET AL.}
\titlerunninghead{EMPIRE solar irradiance reconstruction}
\authoraddr{Corresponding author: K.~L.~Yeo, Max-Planck-Institut f\"{u}r Sonnensystemforschung, Justus-von-Liebig-Weg 3, 37077 G\"{o}ttingen, Germany. (yeo@mps.mpg.de)}

\begin{document}

\title{EMPIRE: A robust empirical reconstruction of solar irradiance variability}
\author{K.~L.~Yeo,\altaffilmark{1} N.~A.~Krivova,\altaffilmark{1} S.~K.~Solanki,\altaffilmark{1,2}}
\altaffiltext{1}{Max-Planck-Institut f\"{u}r Sonnensystemforschung, G\"{o}ttingen, Germany.}
\altaffiltext{2}{School of Space Research, Kyung Hee University, Yongin, Gyeonggi, Korea.}
\begin{abstract}
We present a new empirical model of {total and spectral solar irradiance (TSI and SSI)} variability entitled EMPirical Irradiance REconstruction (EMPIRE). As with existing empirical models, TSI and SSI variability is given by the linear combination of solar activity indices. In empirical models, UV SSI variability is usually determined by fitting the rotational variability in activity indices to that in measurements. Such models have to date relied on ordinary least squares regression, which ignores the uncertainty in the activity indices. In an advance from earlier efforts, the uncertainty in the activity indices is accounted for in EMPIRE by the application of an error-in-variables regression scheme, making the resultant UV SSI variability more robust. The result is consistent with observations and unprecedentedly, with that from other modelling approaches, resolving the long-standing controversy between existing empirical models and other types of models. We demonstrate that earlier empirical models, by neglecting the uncertainty in activity indices, underestimate UV SSI variability. The reconstruction of TSI and visible and IR SSI from EMPIRE is also shown to be consistent with observations. The EMPIRE reconstruction is of utility to climate studies as a more robust alternative to earlier empirical reconstructions. 
\end{abstract}

\begin{article}

\section{Introduction}
\label{introduction}

Solar radiative forcing is a key input to climate models \citep{haigh07,gray10,jungclaus10,schmidt11,matthes16}. It is described in terms of total and spectral solar irradiance, TSI and SSI. TSI is the wavelength-integrated and SSI the per unit wavelength Earthward solar radiative flux (i.e., energy per unit time and area) at the mean Sun-Earth separation. TSI and ultraviolet (UV) SSI have been monitored from space, almost without interruption, since 1978 through a succession of satellite missions \citep{willson81,rottman88,floyd03,deland08,frohlich12,ermolli13,kopp14}. Regular space-based measurement of visible and infrared (IR) SSI started more recently, in 2003, with the SIM instrument onboard the SORCE mission \citep{harder05a,harder05b,harder09}.

The 11-year solar activity cycle sees the cyclic emergence and evolution of kiloGauss-strength magnetic concentrations on the solar surface, which manifests as bright network and faculae and dark sunspots and pores. The satellite observation of solar irradiance reveal it to vary along with this 11-year cycle \citep{willson81,hudson82} and also to rise/fall as solar rotation brings bright/dark magnetic structures across the Earth-facing side of the Sun \citep{willson88,hickey88}. Given the apparent correlation to solar surface magnetism, the surveillance of solar irradiance from space has been accompanied by the development of models seeking to reproduce the variability by modelling it as the sum effect of facular brightening and sunspot darkening. Early examples include \cite{hudson82}, \cite{oster82}, \cite{chapman86} and \cite{foukal86}. Such models have seen a certain degree of success in replicating observed TSI and SSI variability, lending credence to the notion that solar irradiance variability is at least strongly related to solar surface magnetism \citep{domingo09,solanki13}.

The body of solar irradiance measurements extend a period of less than five decades and even then with gaps in time and wavelength coverage. Moreover, due to the challenge in instrument calibration, reconciling the differences between the records from the various monitoring missions remains intractable \citep{deland08,frohlich12,ermolli13,kopp14}. As a result, apart from aiding our understanding of the physical processes driving the variability, solar irradiance models have also risen to provide the extended, uninterrupted and coherent TSI and SSI time series required by climate models.

Solar irradiance models fall into two broad categories according to the approach taken to determine facular brightening and sunspot darkening from solar observations, termed proxy \citep[e.g.,][]{lean88,chapman96,chapman13,lean97,morrill11,thuillier12,coddington16} and semi-empirical \citep[e.g.,][]{fligge00,ermolli03,krivova03,krivova06,fontenla11,shapiro11,ball12,bolduc12,yeo14b}.

In the proxy approach, solar irradiance variability is given by the linear combination of indicators of solar activity acting as proxies of facular brightening and sunspot darkening. Facular brightening is usually represented by the Mg II index \citep{heath86} and sunspot darkening by the Photometric Sunspot Index or PSI \citep{hudson82,frohlich94}. The coefficients of the linear combination is determined from the regression of the activity indices to measured solar irradiance. In the semi-empirical approach, solar surface coverage by magnetic features and its evolution with time is taken directly from spatially-resolved full-disc observations or inferred from activity indices. The intensity spectrum of solar surface features is calculated from models of their atmospheric structures with radiative transfer codes. Solar irradiance at a given time is recovered by assigning the calculated intensity spectra to each position on the solar disc according to the surface coverage and integrating the result over the solar disc.

Proxy and semi-empirical reconstructions of solar irradiance differ in terms of the secular trend and the wavelength-dependence of the variability \citep{ermolli13,solanki13,yeo14a,yeo16}. Critically, proxy and semi-empirical models diverge on the amplitude of the variation over the solar cycle in the UV, an important spectral region for Sun-climate interactions \citep{haigh94,haigh07,gray10}.

The proxy model NRLSSI \citep{lean97,lean00a,coddington16} and the semi-empirical model SATIRE-S \citep{unruh99,fligge00,krivova03,krivova06,ball12,yeo14b} are the two most commonly employed in climate simulations. Over 240 to 400 nm, solar cycle variability in NRLSSI is only about half that in SATIRE-S \citep[see Fig. 7 in][]{yeo14a}. Consequently, their application to climate models results in non-trivial differences in the response in stratospheric ozone, temperature and heating rates \citep[see, for example,][]{oberlander12,ermolli13,dhomse13,ball14,ball16}.

This discrepancy is not confined to just between NRLSSI and SATIRE-S. NRLSSI is based on the regression of the variability at solar rotational timescales in the Mg II index and PSI to that in the UV SSI record from UARS/SOLSTICE \citep{rottman01}, thus assuming that the apparent relationship between the activity indices and measured SSI at rotational timescales applies at all timescales. \cite{thuillier14} noted that UV SSI solar cycle variability in the MGNM model \citep{thuillier12}, which takes the same approach as NRLSSI except it is based on SBUV/2 SSI longwards of 170 nm \citep{deland93}, is similarly weaker than not just SATIRE-S but also the semi-empirical reconstructions by \cite{shapiro11} and \cite{bolduc12} \cite[see table 6 in][]{thuillier14}. Also, UV SSI solar cycle variability in the proxy model by \cite{morrill11}, which departs from NRLSSI and MGNM in that it is based on the direct regression of the Mg II index to measured SSI \citep[in this case, the record from UARS/SUSIM,][]{brueckner93,floyd03} is in close agreement with SATIRE-S \citep{yeo14a,yeo14b,yeo15}. In this paper, we will refer to the \cite{morrill11} model as Mea11.

The variation in UV SSI over the solar cycle declines rapidly with wavelength, from tens of percent at the Lyman-$\alpha$ line to less than a percent longwards of 300 nm. As a result, going towards longer wavelengths, solar cycle variability is increasingly obscured in direct observations by measurement uncertainty \citep[see Fig. 4 in][]{yeo15}. This is the reason why proxy models of UV SSI, with the exception of Mea11, take to comparing just the rotational variability in activity indices to that in measured SSI. \cite{lean00a} stated that in taking the apparent relationship between the activity indices and measured SSI at rotational timescales to apply at all timescales, the recovered solar cycle variability represents the lower limit, a point reiterated by \cite{ermolli13} and \cite{coddington16}. In other words, solar cycle variability might be underestimated from ignoring possible differences in how activity indices relate to solar irradiance at rotational and at solar cycle timescales. This prompted some of the changes introduced to NRLSSI by \cite{coddington16}, which we will examine in Sect. \ref{nrlssi}. Hereafter, we will refer to the original and the updated version of NRLSSI as NRLSSI1 and NRLSSI2, respectively.

It is not certain why the UV SSI solar cycle variability produced by proxy models based on measured rotational variability (NRLSSI1 and MGNM) is weaker than that by other models, both proxy (Mea11) and semi-empirical (SATIRE-S, \citealt{shapiro11} and \citealt{bolduc12}).

Proxy models have, to date, relied on ordinary least squares (OLS) regression when comparing activity indices to measured solar irradiance. The formulation of the OLS assumes the uncertainty in the predictors, the activity indices in this context, is negligible. It is known that uncertainty in the predictors can bias the regression coefficients from OLS towards zero. This is a well-established property of OLS termed regression attenuation \citep{spearman04}. If the uncertainty in the rotational variability in the activity indices used in proxy models is severe enough to effect regression attenuation in the OLS fitting to measured UV SSI rotational variability, the regression coefficients, and therefore reconstructed UV SSI variability, will be underestimated.

In this study, we aim to address this issue by investigating if the relatively weak UV SSI solar cycle variability produced by proxy models such as NRLSSI1 and MGNM is due to differences in how activity indices relate to solar irradiance at rotational and solar cycle timescales \citep[as asserted by][]{lean00a,ermolli13,coddington16} or a result of regression attenuation. To this end, we present a new proxy model of TSI and SSI, which we term EMPirical Irradiance REconstruction (EMPIRE). The model is similar to existing proxy models, except that UV SSI variability is given not by the OLS regression but by the orthogonal distance regression \citep[ODR,][]{fuller87} of the rotational variability in activity indices to that in measured SSI. The ODR is an example of what are termed error-in-variables regression schemes, which are designed to circumvent regression attenuation by taking the uncertainty in the predictors into account.

In the following, we present the EMPIRE model in detail (Sect. \ref{empire}) before discussing the differences from establishing UV SSI variability via ODR instead of OLS (Sect. \ref{odrversusols}). We compare the EMPIRE reconstruction of TSI and SSI to measurements and other models in Sect. \ref{comparison} before a summary and concluding remarks in Sect. \ref{summary}.

\section{EMPIRE reconstruction of TSI and SSI}
\label{empire}

\subsection{Model}
\label{model}

The EMPirical Irradiance REconstruction (EMPIRE) follows the general approach taken by proxy models of solar irradiance. That is, solar irradiance variability is given by the linear combination of solar activity indices acting as proxies of facular brightening and sunspot darkening. Model TSI at time $t$, $T_{\rm mod}\left(t\right)$, is defined as
\begin{equation}
	T_{\rm mod}\left(t\right)=K_{F}{}F\left(t\right)+K_{S}{}S\left(t\right)+K_{\rm ref},
\label{tsieqn}
\end{equation}
where $F$ and $S$ denote the activity indices serving as proxies of facular brightening and sunspot darkening, respectively. The proxy time series are scaled by the factors $K_{F}$ and $K_{S}$ and the sum, $K_{F}{}F+K_{S}{}S$, is offset by $K_{\rm ref}$ to set the reconstructed variability on the absolute level of a reference TSI record to yield TSI. Similarly, SSI at wavelength $\lambda$, $I_{\rm mod}\left(\lambda,t\right)$ is given by
\begin{equation}
	I_{\rm mod}\left(\lambda,t\right)=k_{F}\left(\lambda\right)F\left(t\right)+k_{S}\left(\lambda\right)S\left(t\right)+k_{\rm ref}\left(\lambda\right),
\label{ssieqn}
\end{equation}
where $k_{F}$ and $k_{S}$ give the scaling of the proxy time series and $k_{\rm ref}$ the offset that sets the reconstructed variability on the absolute level of a reference solar spectrum. The wavelength range of the model is 115 to 170000 nm.

Next, we describe the activity indices employed as $F$ and $S$ (Sect. \ref{proxy}), and how the coefficients of the model, i.e., the constants in Equations \ref{tsieqn} and \ref{ssieqn}, were determined (Sects. \ref{tsivar} and \ref{ssivar}). This is followed by a discussion of the uncertainty in the reconstructed solar irradiance variability in Sect. \ref{error}.

\subsection{Proxies of facular brightening and sunspot darkening}
\label{proxy}

We employed the Mg II index composite by IUP\footnote{{The Institut f\"ur Umweltphysik (Institute for Environmental Physics) at the University of Bremen.}} and the PSI composite by \cite{balmaceda09} as proxies of facular brightening and sunspot darkening, respectively ($F$ and $S$ terms in Equations \ref{tsieqn} and \ref{ssieqn}). The IUP Mg II index composite is updated daily. We extended the \cite{balmaceda09} time series to the present with the latest SOON\footnote{The United States Air Force Solar Observing Optical Network.} sunspot area observations.

While the PSI composite extends back to 9 May 1874, the Mg II index composite only goes back to 7 November 1978. To extend the Mg II index composite further back in time, we fit the F10.7 record \citep{tapping87,tapping13}, which started in 14 February 1947, to the Mg II index composite and extended the latter to this date with the result.

There are gaps in the F10.7 (634 days between 15 February 1947 and 29 October 1972) and PSI time series (about seven days per year on average). We interpolated over the gaps in the F10.7 record, the longest of which is six days. We filled the gaps in the PSI composite with the regression of the international sunspot number \citep[version 2,][]{clette16} to the PSI composite.

With these steps, the proxy time series, and therefore the reconstruction, extends from 14 February 1947 to the present at daily cadence with no interruptions. (At the time of writing, the SOON sunspot area record, and therefore the EMPIRE reconstruction, goes up to 30 September 2016.) We note here that over this period, versions 1 and 2 of the international sunspot number differ mainly by the absolute scale such that there is essentially no difference to employing either version to fill the gaps in the PSI composite.

\subsection{TSI reconstruction}
\label{tsivar}

The coefficients of the TSI model ($K$ terms in Equation \ref{tsieqn}) were fixed by the OLS regression of the proxy time series to the SORCE/TIM TSI record \citep[version 17,][]{kopp05c}. We constrained the regression such that the reconstruction matches the TIM record at the December 2008 solar cycle minimum. Here and in the rest of the study, we take the timing of solar cycle minima and maxima established by NOAA\footnote{The National Oceanic and Atmospheric Administration. See www.ngdc.noaa.gov/stp/space-weather/solar-data/solar-indices/sunspot-numbers/cycle-data/.}.

\begin{figure}
\includegraphics[width=\textwidth]{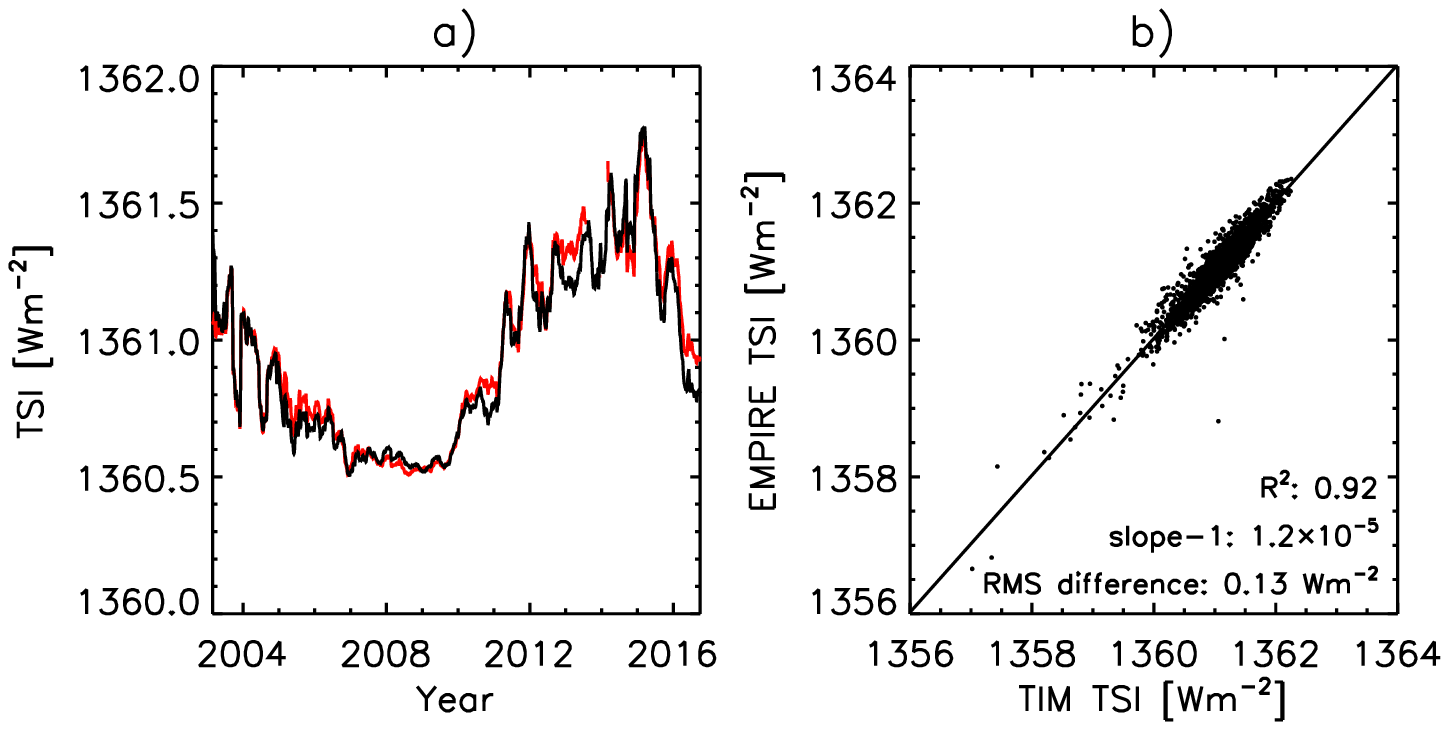}
\caption{a) 81-day moving average of EMPIRE (black) and TIM TSI (red). b) Scatter plot of EMPIRE and TIM TSI (the actual values, not the 81-day moving average) and the corresponding linear fit. The correlation, $R^2$, and RMS difference between the two data sets, and the deviation in the slope of the fit from unity are annotated.}
\label{empssivstim}
\end{figure}

The regression of the proxy time series to the TIM record is robust, as visibly evident in Fig. \ref{empssivstim}. This is also apparent in the correlation ($R^2=0.92$), the root-mean-square or RMS difference (0.13 ${\rm Wm^{-2}}$) and how close the slope of the linear fit to the scatter plot of EMPIRE and TIM TSI is to unity ($1.2\times10^{-5}$).

\subsection{SSI reconstruction}
\label{ssivar}

\subsubsection{UV SSI variability (115 to 420 nm)}
\label{uvssivar}

SSI variability is characterized by $k_{F}$ and $k_{S}$ (Equation \ref{ssieqn}). For the wavelength range of 115 to 420 nm, we determined $k_{F}$ and $k_{S}$ from the regression of the rotational variability in the proxy time series to that in SSI observations, as typical of proxy models. We made use of the SSI records from the SUSIM and SOLSTICE instruments onboard the UARS mission, and SIM and SOLSTICE onboard SORCE, described in Table \ref{ssirecords}. In a departure from earlier proxy models \citep[such as,][]{lean97,thuillier12,coddington16}, in the regression, we execute ODR instead of OLS.

Let $F^{\rm rot}$, $S^{\rm rot}$ and $I^{\rm rot}_{\rm obs}\left(\lambda\right)$ denote the rotational variability in $F$, $S$ and observed SSI, respectively, isolated by subtracting from each time series the corresponding 81-day moving average. The regression of $F^{\rm rot}$ and $S^{\rm rot}$ to $I^{\rm rot}_{\rm obs}\left(\lambda\right)$ is equivalent to finding the best fit plane to the 3D scatter plot of the three variables. For a given variable $x$, let $\epsilon_x$ represent the uncertainty and $\Delta_{i,x}$ the difference between the $i$-th data point (in a total of $n$) and the fit in the $x$-direction. OLS and ODR differ in how the scatter between data and fit, which is minimized in the regression, is defined. While OLS returns the plane that minimizes
\begin{equation}
	\sum^{n}_{i=1}\Delta_{i,I^{\rm rot}_{\rm obs}\left(\lambda\right)}^2,
\label{olseqn}
\end{equation}
ODR seeks the plane that minimizes
\begin{equation}
	\sum^{n}_{i=1}\left[\Delta_{i,I^{\rm rot}_{\rm obs}\left(\lambda\right)}^2+\frac{\epsilon_{F^{\rm rot}}^2}{\epsilon_{I^{\rm rot}_{\rm obs}\left(\lambda\right)}^2}\Delta_{i,F^{\rm rot}}^2+\frac{\epsilon_{S^{\rm rot}}^2}{\epsilon_{I^{\rm rot}_{\rm obs}\left(\lambda\right)}^2}\Delta_{i,S^{\rm rot}}^2\right],
\label{odreqn}
\end{equation}
which is equivalent to minimizing the variance-weighted orthogonal distance between the 3D scatter plot and the plane. Equation \ref{odreqn} reduces to Equation \ref{olseqn} when the variance in the predictors, $\epsilon_{F^{\rm rot}}$ and $\epsilon_{S^{\rm rot}}$, are indeed negligible as assumed in OLS.

Fitting $F^{\rm rot}$ and $S^{\rm rot}$ to $I^{\rm rot}_{\rm obs}$ via ODR requires \textit{a priori} knowledge of $\epsilon_{F^{\rm rot}}$, $\epsilon_{S^{\rm rot}}$ and $\epsilon_{I^{\rm rot}_{\rm obs}}$ (Equation \ref{odreqn}). We estimated $\epsilon_{F^{\rm rot}}$ by comparing the IUP Mg II index composite to the competing composite by LASP\footnote{The Laboratory for Atmospheric and Space Physics at the University of Colorado Boulder.} \citep{
snow05b} and $\epsilon_{S^{\rm rot}}$ by considering the uncertainty in the sunspot area measurements underlying the PSI, detailed in Appendix \ref{efrotesrot}. The signal-to-noise ratio or S/N of $F^{\rm rot}$ and $S^{\rm rot}$, given by the ratio of signal variance to noise variance, is 12.5 and 12.4, respectively.

For each SSI record, for each wavelength bin between 115 and 420 nm, we estimated $\epsilon_{I^{\rm rot}_{\rm obs}}$ along with $k_{F}$ and $k_{S}$ by the following procedure.
\begin{enumerate}
	\item We regressed $F^{\rm rot}$ and $S^{\rm rot}$ to $I^{\rm rot}_{\rm obs}$ via OLS.
	\item We determined $\epsilon_{I^{\rm rot}_{\rm obs}}$ as a function of time. For each day, $\epsilon_{I^{\rm rot}_{\rm obs}}$ is given by the RMS difference between $I^{\rm rot}_{\rm obs}$ and the fit within a 731-day window centred on that day.
	\item Taking the estimate of $\epsilon_{I^{\rm rot}_{\rm obs}}$ from the step 2, we regressed $F^{\rm rot}$ and $S^{\rm rot}$ to $I^{\rm rot}_{\rm obs}$ via ODR. We made use of the implementation of ODR in the Python SciPy package \citep{boggs89}.
	\item We repeated steps 2 and 3, updating $\epsilon_{I^{\rm rot}_{\rm obs}}$ and the fit, till the regression coefficients stabilize. The final fit is self-consistent in that the estimate of $\epsilon_{I^{\rm rot}_{\rm obs}}$ used in the ODR fitting (step 3) is similar to the $\epsilon_{I^{\rm rot}_{\rm obs}}$ calculated with the fit (step 2).
\end{enumerate}
For the wavelength bins below 166 nm, we fixed $k_{S}$ at null. The effect on reconstructed UV SSI variability is negligible. It is well-established that the contribution by sunspot darkening to solar irradiance variability is minute in the far UV (FUV) or below about 200 nm \citep{unruh08}. {Here, faculae are much brighter than the quiet Sun than sunspots are dark such that solar irradiance variability is dominated by the former.} Indeed, proxy models often neglect sunspot darkening in the FUV \citep[for example,][]{woods15} or across the entire UV \citep{morrill11,thuillier12}.

\begin{figure}
\includegraphics[width=\textwidth]{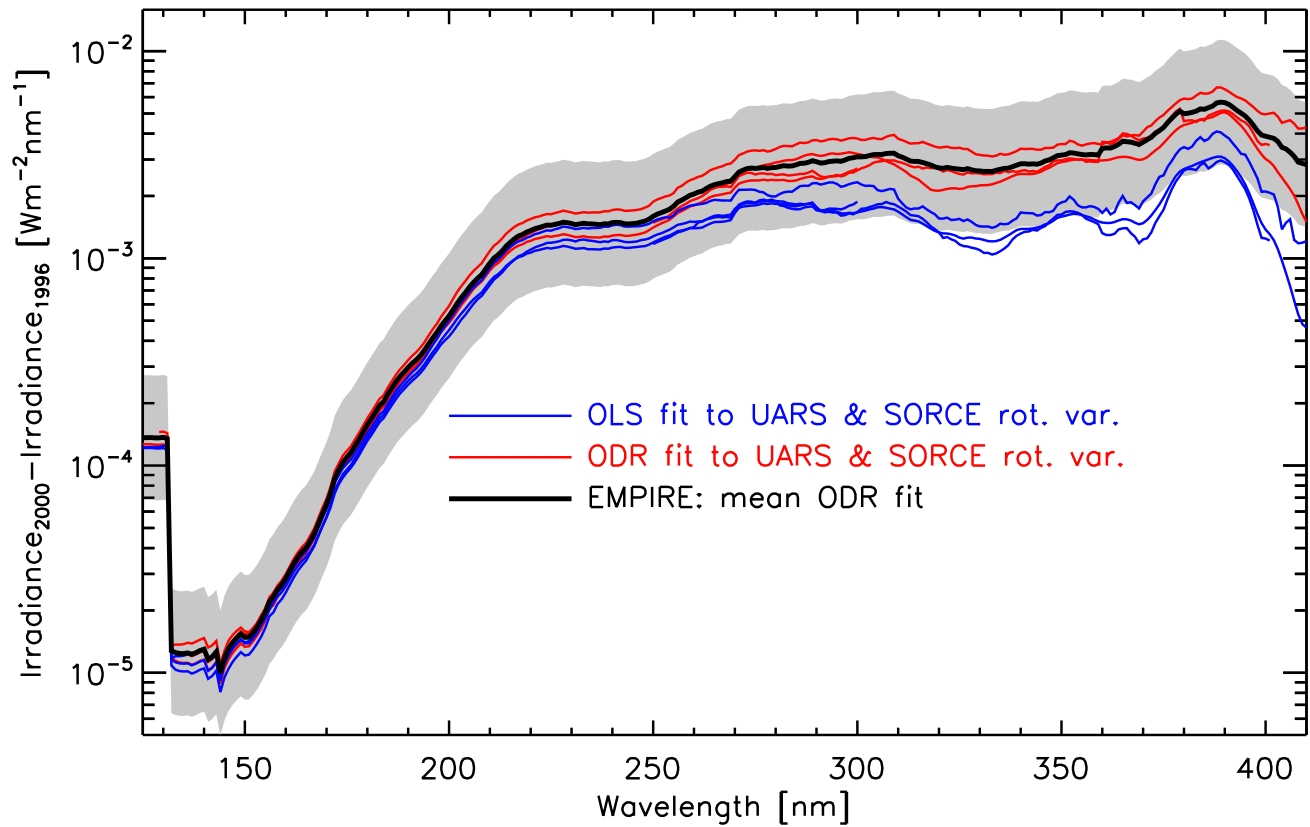}
\caption{As a function of wavelength, the change in UV SSI between the 1996 solar cycle minimum and the 2000 maximum. The 20 nm moving average is depicted. The values inferred from the regression of the rotational variability in the proxy time series to that in each of the UARS and SORCE SSI records (see Table \ref{ssirecords}) via ODR and OLS are drawn in red and blue, respectively. There are four SSI records, giving the four estimates for each regression method. The black line denotes EMPIRE, which is based on the average of the results of the ODR analysis of the various SSI records, and the shaded region the corresponding 50\% and 200\% bounds. See Sect. \ref{uvssivar} for details.}
\label{empssisingleestimates}
\end{figure}

In Fig. \ref{empssisingleestimates}, we chart the variation in UV SSI over the ascending phase of solar cycle 23 estimated from the ODR analysis of each SSI record (red). As evident, the results from the various records are consistent with one another. For the final reconstruction, we took the mean of the $k_{F}$ and $k_{S}$ determined from the various records. The resultant UV SSI variability is drawn in black. To aid the reader, a summary of the UV SSI reconstructions depicted in Fig. \ref{empssisingleestimates} and to be examined in a similar manner later in this paper (Figs. \ref{empssiodrvsols} and \ref{empssivsnrlssi}) is given in Table \ref{reconstructiontable}.

UV SSI variability inferred from the ODR analysis of the various SSI records (red, Fig. \ref{empssisingleestimates}), the mean of which is taken into EMPIRE (black), is on the whole greater than that from applying OLS instead (blue). The difference is minute in the FUV but increases markedly with wavelength such that longwards of 300 nm, the variability from the ODR analysis is about twice that from OLS. We will examine the possible causes of this divergence between the ODR and OLS analyses of observed SSI rotational variability, including regression attenuation, in Sect. \ref{odrversusols}.

\subsubsection{Visible and IR SSI variability (420 to 170000 nm)}
\label{visirssivar}

We determined SSI variability in the visible and IR (420 to 170000 nm) from the calculated intensity of solar surface features. This follows the NRLSSI1 (400 to 100000 nm segment) and NRLSSI2 (2400 to 100000 nm) reconstructions. Except, while NRLSSI made use of the intensity contrast of faculae and sunspots calculated by \cite{solanki98}, we employed the intensity spectra of quiet Sun, faculae, and sunspot umbra and penumbra from \cite{unruh99}. The \cite{unruh99} study is actually a rigorous update of the \cite{solanki98} work; while the \cite{solanki98} results were based on an empirical model, the \cite{unruh99} intensity spectra came from a proper solution of the radiative transfer equation.

\cite{unruh99} calculated intensity spectra, extending 10 to 170000 nm, of each feature type at heliocentric angles between $0^{\circ}$ to $87^{\circ}$. By the appropriate summation over heliocentric angles, we calculated the SSI if the solar surface is entirely covered by each feature type. Let $I_{\rm qsn}$, $I_{\rm fac}$, $I_{\rm umb}$ and $I_{\rm pen}$ denote the result for quiet Sun, faculae, umbra and penumbra, respectively. Taking into account that the observed umbral to penumbral area ratio of sunspots is about 1:4 on average \citep{solanki03,wenzler05}, the SSI when the solar surface is entirely covered by sunspots, $I_{\rm spt}$ is given by $0.2I_{\rm umb}+0.8I_{\rm pen}$.

If faculae are evenly distributed on the solar surface, then $k_{F}\left(\lambda\right)$ would be proportional to $I_{\rm fac}\left(\lambda\right)-I_{\rm qsn}\left(\lambda\right)$, and likewise for sunspots. Taking this assumption, we define
\begin{equation}
	k_{F}\left(\lambda\right)=f_{F}\left[I_{\rm fac}\left(\lambda\right)-I_{\rm qsn}\left(\lambda\right)\right]
\label{fbeqn}
\end{equation}
and
\begin{equation}
	k_{S}\left(\lambda\right)=f_{\rm S}\left[I_{\rm spt}\left(\lambda\right)-I_{\rm qsn}\left(\lambda\right)\right],
\label{sdeqn}
\end{equation}
where $f_{F}$ and $f_{S}$ are constants, in the visible and IR. The wavelength range of EMPIRE (115 to 170000 nm) encompasses the bulk of the energy in solar radiative flux such that we can require wavelength-integrated SSI to be equal to TSI in the reconstruction. Taking this into account, $f_{F}$ is fixed by imposing that the integral of $k_{F}$ over 115 to 170000 nm equates to $K_{F}$, and likewise for $f_{S}$.

Faculae and sunspots are of course not evenly distributed on the solar surface. However, this is a necessary assumption as EMPIRE, as with all other proxy models, relies on Sun-as-a-star measures of solar activity, where there is no information about the spatial distribution of faculae and sunspots. We will demonstrate in Sect. \ref{visircompare} that the uncertainty in visible and IR SSI variability from this assumption is minute. It is worth pointing out that the process here is an improvement over that in NRLSSI. In NRLSSI, SSI variability is estimated from expressions similar to Equations \ref{fbeqn} and \ref{sdeqn} based on the \cite{solanki98} calculations, which not only neglects the non-uniform distribution of faculae and sunspots, but also the variation in their intensity contrast with heliocentric angle.

\subsubsection{Reference spectrum}

Finally, reconstructed SSI variability is set onto a reference spectrum to yield SSI. The 115 to 2400 nm segment of the reference spectrum is given by the Whole Heliosphere Interval or WHI \citep{thompson11} quiet Sun reference spectrum \citep{woods09}, essentially the mean SSI in the period of 10 to 16 April 2008. The 2400 to 170000 nm segment is given by $I_{\rm qsn}$ (Sect. \ref{visirssivar}), scaled such that the integral of the reference spectrum over 115 to 170000 nm is equal to the mean reconstructed TSI (Sect. \ref{tsivar}) over the period of 10 to 16 April 2008. The $k_{\rm ref}$ term in the SSI model (Equation \ref{ssieqn}) is fixed such that the mean reconstructed SSI over the same period matches the reference spectrum. We note that by the steps taken here and in the previous section, the integral of reconstructed SSI over the wavelength range of the model matches reconstructed TSI exactly.

\subsection{Error analysis}
\label{error}

There is systematic uncertainty in the reconstructed solar irradiance variability from assuming the Mg II index and PSI to be appropriate proxies of facular brightening and sunspot darkening over the entire wavelength range of the model and that the apparent relationship between them and measured SSI at rotational timescales applies at longer timescales \citep[see discussion in][]{yeo14b}. This is of course not unique to EMPIRE but an open question about proxy models in general. Quantifying this uncertainty is beyond the scope of the present study. As we will see in Sect. \ref{comparison}, the EMPIRE reconstruction compares well to measurements and certain other models, giving us confidence that this systematic uncertainty is, at least to within the limits of this agreement, weak.

Here, we examine the uncertainty in reconstructed variability resulting from the uncertainty in the $K_{F}$ and $K_{S}$ terms in the TSI model (Equation \ref{tsieqn}) and the $k_{F}$ and $k_{S}$ terms in the SSI model (Equation \ref{ssieqn}). For TSI and UV SSI, where the coefficients are determined from the regression of the proxy time series to measured solar irradiance (Sects. \ref{tsivar} and \ref{uvssivar}), the uncertainty is given by the formal regression error. For visible and IR SSI, the coefficients are constrained by the TSI and UV SSI coefficients and the \cite{unruh99} intensity spectra (Sect. \ref{visirssivar}). Accordingly, the uncertainty in the visible and IR coefficients is propagated from the uncertainty in the TSI and UV SSI coefficients.

In Table \ref{uncertainty}, we list the uncertainty in the change between the 1996 solar cycle minimum and the 2000 maximum in TSI and in the integrated SSI over certain wavelength intervals, expressed as a percentage. The wavelength intervals encompass spectral features relevant to climate studies, namely, the Lyman-$\alpha$ line, the Schuman-Runge oxygen continuum and bands, the Herzberg oxygen continuum, the Hartley and Higgins ozone bands, and water vapour and carbon dioxide bands in the shortwave IR. For the wavelength intervals in the UV, the uncertainty rises steadily with wavelength. This just reflects the fact that UV SSI observations become more and more uncertain with {increasing} wavelength \citep[see][and Sect. \ref{measurements}]{yeo15}.

\section{Why does the ODR and OLS regression of the proxy time series to observed UV SSI rotational variability differ?}
\label{odrversusols}

The key departure in EMPIRE from earlier proxy models based on measured rotational variability, including NRLSSI, is the application of orthogonal distance regression (ODR) instead of ordinary least squares (OLS) when fitting the rotational variability in the proxy time series to that in observed SSI (i.e., $F^{\rm rot}$ and $S^{\rm rot}$ to $I^{\rm rot}_{\rm obs}$). Recall from Sect. \ref{uvssivar}, the variability in UV SSI based on the ODR analysis of the UARS and SORCE records (red, Fig. \ref{empssisingleestimates}) is stronger than that from the OLS analysis (blue) and increasingly so with wavelength. EMPIRE UV SSI variability, drawn in black, is essentially the average of the estimates from the ODR analysis of the various SSI records. The discrepancy between the ODR and OLS analyses could have arisen from the following.
\begin{enumerate}
	\item The uncertainty in the proxy time series could have biased the OLS regression coefficients towards zero (i.e., regression attenuation).
	\item As suggested by \cite{yeo15}, the solar cycle variability inferred from such an approach could be downward biased by the uncertainty in measured SSI rotational variability.
	\item The ODR fitting requires the uncertainty in the proxy and SSI time series as input (Equation \ref{odreqn}) and could therefore be biased by their misestimation.
\end{enumerate}
To determine which of these is responsible for the divergence between the ODR and OLS analyses, we conducted the tests described next in Sects. \ref{cleaniobsrot} and \ref{cleanfrotsrot}.

\subsection{Influence of the uncertainty in observed SSI}
\label{cleaniobsrot}

To elucidate the influence of the uncertainty in observed SSI on the ODR and OLS analyses, we denoised $I^{\rm rot}_{\rm obs}$ and repeated the analyses on the result.

We denoised $I^{\rm rot}_{\rm obs}$ by a procedure based on the non-local means filter or NLMF \citep{baudes05}. The signal on a given day is replaced by the average of the time series, weighted such that the bulk of the weight comes from the days with similar rotational variability in solar activity. The denoising scheme is detailed in Appendix \ref{nlmf}.

In Fig. \ref{empssinlmf}, we depict $I^{\rm rot}_{\rm obs}$ at 129.5 nm and 379.5 nm over 1994 in the SUSIM record, before (blue) and after denoising (red). Let us take the EMPIRE reconstruction, drawn in black, as a gauge of the underlying signal (we will demonstrate the robustness of EMPIRE rotational variability in Sect. \ref{uvcompare}). The effectiveness of the denoising scheme is evident in how the SUSIM values compare to EMPIRE before and after denoising.

In Fig. \ref{empssiodrvsols}a, we compare the UV SSI variability inferred from repeating the ODR and OLS analyses on denoised SUSIM $I^{\rm rot}_{\rm obs}$ (dashed lines) along the original results (solid lines). For both ODR (red) and OLS (blue), denoising $I^{\rm rot}_{\rm obs}$ made no appreciable difference. The same is noted of the UARS/SOLSTICE, SIM and SORCE/SOLSTICE records, not shown to avoid repetition. As an additional test, we calculated the mean SSI over wavelength intervals between 3 and 7 nm, and repeated the ODR and OLS analyses on the result. Again, the denoising that this spectral averaging effected had almost no effect. {Denoising measured SSI rotational variability made no appreciable difference to the ODR and OLS analyses; above 300 nm, the UV SSI variability inferred from the ODR analysis remains about twice that from the OLS analysis (Fig. \ref{empssiodrvsols}a). This rules out the uncertainty in observed SSI as the main cause of the divergence between the two.}

\subsection{Influence of the uncertainty in the proxy time series}
\label{cleanfrotsrot}

Next, we examined the influence of the uncertainty in the proxy time series on the OLS analysis. To this end, we identified 13-day periods where rotational variability is dominated by the transit of active region(s) across the solar disc, termed here as transit periods. Taking the $F^{\rm rot}$, $S^{\rm rot}$ and $I^{\rm rot}_{\rm obs}$ time series, we averaged the data from multiple transit periods to yield mean transit profiles and repeated the OLS analysis with the result. The idea here is to derive and repeat the OLS analysis on denoised proxies of $I^{\rm rot}_{\rm obs}$, $F^{\rm rot}$ and $S^{\rm rot}$.

The transit periods are identified as follows. We smoothed $F^{\rm rot}$ and $S^{\rm rot}$ with a binomial filter \citep{marchand83} and identified as transit periods 13-day intervals where $F^{\rm rot}$ ascends monotonically up to the 7th day before descending steadily to the 13th day and $S^{\rm rot}$ is higher, on average, over the 5th to 9th day than over the 1st to 3rd and 11th to 13th day. Let $N$ represent the number of transit periods, for a given SSI record, for which there is complete data; $N=7$ for SUSIM, 8 for UARS/SOLSTICE, 41 for SORCE/SOLSTICE and 42 for SORCE/SIM.

At each wavelength bin, we did the following. Taking each record, we averaged the 13-day $I^{\rm rot}_{\rm obs}$ segments from the $N$ transit periods, giving us a $I^{\rm rot}_{\rm obs}$ mean transit profile. We took the $N$-weighted average of the $I^{\rm rot}_{\rm obs}$ mean transit profiles from the various records, yielding one final $I^{\rm rot}_{\rm obs}$ mean transit profile. The $F^{\rm rot}$ and $S^{\rm rot}$ time series are analogously sampled to yield $F^{\rm rot}$ and $S^{\rm rot}$ mean transit profiles corresponding to the $I^{\rm rot}_{\rm obs}$ mean transit profile.

In Fig. \ref{empssiodrvsols}b, we see that the UV SSI variability inferred from the OLS regression of the $F^{\rm rot}$ and $S^{\rm rot}$ mean transit profiles to the $I^{\rm rot}_{\rm obs}$ mean transit profiles (blue dotted line) aligns closely with EMPIRE (black). That is to say, for OLS, considering the denoised proxies of $I^{\rm rot}_{\rm obs}$, $F^{\rm rot}$ and $S^{\rm rot}$ instead of the original time series yields results close to what we get from the ODR analysis of the original time series. Given denoising $I^{\rm rot}_{\rm obs}$ has no appreciable effect on the OLS analysis, demonstrated in the previous section, we attribute the effect on the OLS analysis to the denoising of $F^{\rm rot}$ and $S^{\rm rot}$. The results of this test indicate the following.
\begin{enumerate}
	\item The divergence between the ODR and OLS analyses in Sect. \ref{uvssivar} is mainly from regression attenuation bias in the OLS case. {If regression attenuation had not occurred in the OLS analysis of the original data, the results should be similar to that from the OLS analysis of the mean transit profiles.}
	\item	The uncertainty in the ODR analysis, and therefore EMPIRE, from any misestimation of the uncertainty in the proxy and SSI time series is likely minute. {If the uncertainty in any of the variables is misestimated, the ODR analysis and the OLS analysis of the mean transit profiles would diverge.}
\end{enumerate}
The OLS regression of the rotational variability in the proxy time series to that in measured SSI sees an underestimation of SSI variability due to regression attenuation. We saw in Fig. \ref{empssivstim} that the OLS regression of the proxy time series to the TIM record, which underpins the TSI reconstruction, does not suffer regression attenuation. The S/N of $F^{\rm rot}$ and $S^{\rm rot}$ is lower than that of $F$ and $S$ due to the variability removed when isolating the rotational variability. This is why the OLS regression of the rotational variability in the proxy time series to that in measured SSI is susceptible to regression attenuation while the OLS regression of the proxy time series to measured TSI is not.

The observation that the discrepancy between the ODR and OLS analyses widens with wavelength (Fig. \ref{empssisingleestimates}), we surmise, arises from the wavelength-dependence of the relative influence of sunspot darkening and facular brightening \citep{unruh08}. As sunspot darkening is negligible in the FUV (Sect. \ref{uvssivar}), the uncertainty in the PSI has minimal influence on the OLS regression here. Towards longer wavelengths, as sunspot darkening strengthens in relation to facular brightening, so does the influence of the uncertainty in the PSI on the OLS regression.

\section{Comparison with measurements and other models}
\label{comparison}

In this section, we verify the EMPIRE reconstruction against measurements and other models of solar irradiance. We will discuss the variability in UV SSI (Sect. \ref{uvcompare}), in visible and IR SSI (Sect. \ref{visircompare}), and in TSI (Sect. \ref{tsicompare}) in turn.

\subsection{UV SSI}
\label{uvcompare}

\subsubsection{Measurements}
\label{measurements}

We compared the EMPIRE reconstruction to the available UV SSI records that extend for at least a few years at daily cadence, cover a solar cycle minimum, and are calibrated for instrument degradation. In Fig. \ref{empssivsuvssi}, we depict the integrated SSI over the wavelength intervals of 120 to 180 nm, 180 to 240 nm, 240 to 300 nm, and 300 to 400 nm. The wavelength intervals are chosen as such to accommodate the different wavelength range of the various records. We smoothed each time series with a 81-day boxcar filter to elucidate the trend over the solar cycle. Concurrent measurements from different instruments can be offset from one another due to radiometric uncertainty. To aid the comparison, we bring the time series from the various records to a common absolute scale by normalizing them to the EMPIRE reconstruction at solar cycle minima.

In the UV, solar cycle variability exhibits a broad decline with wavelength, from a few $10\%$ below 200 nm to less than 1\% above 300 nm \citep[see Fig. 3 in][]{yeo15}. As a consequence, observed solar cycle variability is increasingly obscured by measurement uncertainty.

In the 120 to 180 nm interval (Fig. \ref{empssivsuvssi}a), where apparent solar cycle variability is least affected by measurement uncertainty, the various records are consistent with one another and well-replicated by EMPIRE. A notable exception is the divergence between EMPIRE and TIMED/SEE after 2008. The fact that the 2008 and 2015 levels indicated by the SEE record are similar even though the Sun is more active in 2015 and the agreement between EMPIRE and SORCE/SOLSTICE over the same period both allude to unresolved instrumental effects in the SEE record here.

\cite{yeo15} showed that the long-term uncertainty of Nimbus-7/SBUV, SME, NOAA-9/SBUV2, UARS/SOLSTICE and SIM is, longwards of 200 to 300 nm, severe enough to obscure solar cycle variability. This manifests itself here as these records becoming mutually inconsistent and not varying along with the solar cycle longwards of 180 nm (Figs. \ref{empssivsuvssi}b to \ref{empssivsuvssi}d). As solar cycle variability is increasingly obscured in these records, so the agreement with EMPIRE, where the solar cycle is clearly apparent, deteriorates.

The SUSIM, SORCE/SOLSTICE and Aura/OMI records are stable enough to exhibit solar cycle variability across their respective wavelength range. They are reasonably well-replicated in EMPIRE (Fig. \ref{empssivsuvssi}). Longwards of 180 nm, SORCE/SOLSTICE indicates a steeper decline over 2004 and 2005 than EMPIRE (Figs. \ref{empssivsuvssi}b and \ref{empssivsuvssi}c). EMPIRE does however, reproduce the trend in SUSIM SSI over the same period so this discrepancy could be indicative of unresolved instrumental trends in the SORCE/SOLSTICE record here.

There is to date no reported model that is able to replicate the marked decline in UV SSI between 2003 and 2008 recorded by SIM, and EMPIRE is no exception (Figs. \ref{empssivsuvssi}c and \ref{empssivsuvssi}d). Numerous studies, whether through other measurements, models or theoretical arguments, have concluded that this trend indicated by SIM is not likely solar in origin \citep{ball11,deland12,lean12,unruh12,ermolli13,solanki13,wehrli13,morrill14,yeo14b,yeo14a,woods15}.

Next, let us examine rotation variability (Fig. \ref{empssivsuvssirot}). As with solar cycle variability, rotational variability diminishes with wavelength (note left-hand axis), with the consequence that observed rotational variability is increasingly obscured by measurement uncertainty. In the 120 to 180 nm interval, where observed rotational variability is least affected by measurement uncertainty, the various records are in agreement with one another and well-replicated by EMPIRE. Longwards of 180 nm, as the agreement between the various records deteriorates, reflecting the increasing influence of measurement uncertainty, evidently so does the agreement with EMPIRE. Except at lower wavelengths, it is difficult to discern the rotational variability in UV SSI directly from measurements.

To verify the rotational variability in EMPIRE UV SSI, even at longer wavelengths where observations are obscured by measurement uncertainty, we examined the mean transit profile. As done in Sect. \ref{cleanfrotsrot}, we derived, at each wavelength interval, a mean transit profile combining the transit period data from all of the records examined here apart from Nimbus-7/SBUV, SME and OMI. (There are no transit periods for which there is complete data from the Nimbus-7/SBUV record. We excluded the SME and OMI records on the grounds that the noise is severe enough to detrimentally affect the mean transit profiles.) We sampled the EMPIRE reconstruction in an analogous manner. As indicated by the agreement between the observed and EMPIRE mean transit profiles (Fig. \ref{empssivsmeanprofile}), EMPIRE UV SSI rotational variability is consistent with observations across the wavelength range examined.

In summary, we conclude that the solar cycle and rotational variability in EMPIRE UV SSI is well-supported by available observations.

\subsubsection{NRLSSI1 and NRLSSI2}
\label{nrlssi}

As in EMPIRE, NRLSSI solar irradiance variability is given by the linear combination of the Mg II index and PSI. For the 120 to 400 nm segment of the NRLSSI1 reconstruction, the scaling of the proxy time series is determined from the OLS regression of the rotational variability to that in the UARS/SOLSTICE record \citep{lean97,lean00a}. For the 115 to 2400 nm segment of the NRLSSI2 reconstruction, the scaling is similarly determined from SORCE/SOLSTICE 115 to 309 nm and SIM 309 to 2400 nm SSI \citep{coddington16}. The authors multiplied the scaling coefficients so derived by a set of corrective factors on the assumption that the relationship between the proxy time series and SSI differ at rotational and solar cycle timescales.

In Fig. \ref{empssivsnrlssi}a, we compare the UV SSI variability in the EMPIRE (black), NRLSSI1 ({green} solid line) and NRLSSI2 (violet solid line) reconstructions. They are consistent with one another up to about 200 nm, above which EMPIRE variability is markedly stronger. Looking at the integrated SSI over 300 to 400 nm, EMPIRE variability is 1.97 and 1.48 times that in NRLSSI1 and NRLSSI2, respectively.

We replicated the NRLSSI1 and NRLSSI2 procedure here. The NRLSSI1 reproduction ({green} dashed line, Fig. \ref{empssivsnrlssi}a) is given by the OLS analysis of UARS/SOLSTICE SSI from Sect. \ref{uvssivar}. The NRLSSI2 reproduction (violet dashed line) is similarly from the OLS analysis of the SORCE/SOLSTICE and SIM records. We implemented the same correction introduced by \cite{coddington16}. We also indicate the variability in the NRLSSI2 reproduction if we do not apply this correction (violet dotted line).

The NRLSSI2 reproduction (violet dashed line, Fig. \ref{empssivsnrlssi}a) compares well with NRLSSI2 (violet solid line) while the NRLSSI1 reproduction ({green} dashed line) and the uncorrected NRLSSI2 reproduction (violet dotted line) align closely with NRLSSI1 ({green} solid line). This indicates the following.
\begin{enumerate}
	\item The stronger solar cycle variability in NRLSSI2 as compared to NRLSSI1 results mainly from the correction introduced to NRLSSI2. Without this correction, the variability in NRLSSI1 and NRLSSI2 would have been similar.
	\item Apart from the correction applied to NRLSSI2, the discrepancy between EMPIRE and the two NRLSSI reconstructions is primarily from the execution of ODR in EMPIRE and OLS in NRLSSI.
\end{enumerate}
As established in Sect. \ref{odrversusols}, the OLS regression of the rotational variability in the Mg II index and PSI to that in measured SSI results in an underestimation of solar cycle variability due to regression attenuation. Obviously, UV SSI solar cycle variability is underestimated in the NRLSSI reconstructions.

\subsubsection{Morrill et al. 2011 (Mea11) and SATIRE-S}
\label{othermodels}

Recall from the introduction, longwards of around 240 nm, proxy models based on measured rotational variability yielded weaker solar cycle variability than other models, including Mea11 and SATIRE-S.

Mea11 exploits the fact that the SUSIM record is sufficiently stable to exhibit solar cycle variability over its wavelength range (c.f. Sect. \ref{measurements}), allowing proxy modelling via the direct fitting of proxies of facular brightening and sunspot darkening to the record. Here, we examined the 150 to 411 nm segment of the reconstruction, given by the OLS regression of the Mg II index to the SUSIM record. In employing the Mg II index alone, the assumption is sunspot darkening is negligible over the wavelength range considered. In Fig. \ref{empssivsnrlssi}b, we compare the UV SSI variability in Mea11 to that inferred from the OLS regression of both the Mg II index and PSI to the SUSIM record. As indicated by the agreement, the assumption of negligible sunspot darkening in Mea11 has limited bearing on the solar cycle variability replicated by this model. Regression attenuation is not a concern here. Recall from Sect. \ref{cleanfrotsrot}, the OLS regression of the proxy time series to measured solar irradiance is not susceptible to regression attenuation as when fitting just the rotational variability.

In SATIRE-S, solar irradiance is reconstructed from the same calculated intensity spectra of solar surface features by \cite{unruh99} described in Sect. \ref{visirssivar}. Taking spatially-resolved full-disc intensity images and magnetograms, the solar disc is segmented into quiet Sun, faculae and sunspots. SSI is derived by assigning the appropriate intensity spectrum to each disc position according to the segmentation and the distance from disc centre, and summing the result over the solar disc. Certain empirical corrections, based on the SORCE/SOLSTICE record and the WHI reference spectra, are applied to the 115 to 300 nm segment of the reconstruction. These corrections are such that reconstructed SSI variability longwards of 180 nm remains independent of any observed SSI time series \citep[see][]{yeo14a,yeo15}.

As also noted in the introduction, the discrepancy between proxy models based on measured rotational variability and models such as Mea11 and SATIRE-S has been attributed to the proxy models neglecting differences in how activity indices relate to solar irradiance at rotational and solar cycle timescales. EMPIRE (black, Fig. \ref{empssivsnrlssi}b), which departs from earlier proxy models in that biases from regression attenuation is circumvented by executing ODR instead of OLS, is in agreement with both Mea11 (cyan solid line) and SATIRE-S (orange). It is worth emphasizing here that while EMPIRE is based on the apparent relationship between activity indices and SSI at rotational timescales, Mea11 is based on the apparent relationship at solar cycle timescales, and SATIRE-S SSI variability longwards of 180 nm is independent of any SSI observations. The weaker UV SSI solar cycle variability in proxy models based on measured rotational variability is an artifact of regression attenuation and not, as previously thought, from differences in how activity indices relate to solar irradiance at rotational and at solar cycle timescales. More importantly, the fact that the UV SSI solar cycle variability in EMPIRE is consistent with available observations where they are reliable (Sect. \ref{measurements}) and two completely independent models (Mea11 and SATIRE-S) is strong indication that we are, with these models and measurements, converging on the true level.

\subsection{Visible and IR SSI}
\label{visircompare}

In Fig. \ref{empssispectraldelta}, we compare the change in SSI between 2003, near the start of the SIM record, and the 2008 solar cycle minimum in EMPIRE (black) with that indicated by the SIM, NRLSSI1, NRLSSI2 and SATIRE-S data sets (red). The UV (below 420 nm), discussed in Sects. \ref{uvcompare}, is included for completeness. Here, we focus on the visible and IR (above 420 nm).

The rise in 420 to 1600 nm SSI between 2003 and 2008 recorded by SIM is not reproduced in EMPIRE (Fig. \ref{empssispectraldelta}a) or indeed any other model reported in the literature. This increase, coming at a time when solar activity is declining, conflicts with models of solar irradiance, all of which see SSI vary in phase with the solar cycle in this wavelength range except near 1600 nm (as apparent for the various models depicted in Fig. \ref{empssispectraldelta}). {The only exception is the model presented by \cite{fontenla15} and even then only in a limited wavelength range in the visible (around 450 to 600 nm, see Fig. 12 in their paper).} Just as numerous studies have attributed the acute drop in SIM UV SSI over the same period to unaccounted instrumental effects, mentioned in Sect. \ref{measurements}, various authors have asserted the same of this upward trend in SIM visible and IR SSI \citep{ball11,ermolli13,solanki13,wehrli13,yeo14b,yeo14a,woods15}. It is worth noting that except in the UV, the uncertainty in the change in SIM SSI between 2003 and 2008 (estimated from the reported long-term uncertainty of the record, 0.01\% per year) is actually greater than the recorded variability (yellow shaded region, Fig. \ref{empssispectraldelta}a).

While SIM solar cycle variability is suspect, the rotational variability, up to 1600 nm, is relatively stable and well-replicated by EMPIRE (Figs. \ref{empssivssim}a to \ref{empssivssim}c). Rotational variability in the 1600 to 2416 nm segment of the record, provided by the electrical substitution radiometer in the instrument, is completely hidden by measurement uncertainty (Fig. \ref{empssivssim}d).

EMPIRE visible and IR SSI variability differs markedly from NRLSSI1 (Fig. \ref{empssispectraldelta}b) and NRLSSI2 (Fig. \ref{empssispectraldelta}c), although in the case of NRLSSI2, the trend with wavelength is qualitatively similar.

SSI variability in EMPIRE longwards of 420 nm and in NRLSSI1 longwards of 400 nm is inferred from the calculated intensity of solar surface features. The discrepant variability (Fig. \ref{empssispectraldelta}b) arises from the fact that EMPIRE employs the intensity spectra from \cite{unruh99} and NRLSSI1 the intensity contrasts from \cite{solanki98}, discussed in Sect. \ref{visirssivar}. As pointed out there, while the \cite{unruh99} results came from a proper solution of the radiative transfer equation and were applied in a manner that accounts for the variation in intensity contrast with distance from disc centre, the \cite{solanki98} results were derived empirically and applied with no regard of disc position dependence. EMPIRE is, in this regard, an improvement over NRLSSI1.

SSI variability in NRLSSI2, up to 2400 nm, is determined from the OLS analysis of SORCE SSI rotational variability (Sect. \ref{nrlssi}). The divergence between EMPIRE and NRLSSI2 (Fig. \ref{empssispectraldelta}c) is, at least in part, from the possible influence of regression attenuation and the correction introduced by \cite{coddington16} on NRLSSI2. Apart from the SIM record, the only other daily, extended, instrument degradation corrected record of SSI longwards of 420 nm available is that from OMI, and that only extends up to 500 nm. This makes it challenging, at least within the limits of this study, to verify the effect of regression attenuation in the visible and IR as done for the UV.

SATIRE-S SSI is reconstructed from the same calculated intensity spectra by \cite{unruh99} utilized in EMPIRE. The two models differ in how the intensity spectra are combined. In SATIRE-S, this is based on spatially-resolved full-disc maps of faculae and sunspots (Sect. \ref{othermodels}). In EMPIRE, the fact that we rely on Sun-as-a-star measures of solar activity as proxies of facular and sunspot activity constrained us to combine the intensity spectra in the manner described in Sect. \ref{visirssivar}, which assumes faculae and sunspots are evenly distributed on the solar disc. The fact that SSI variability in EMPIRE and in SATIRE-S remain closely similar (Fig. \ref{empssispectraldelta}d) does suggests that the effect of this assumption is minute.

\subsection{TSI}
\label{tsicompare}

\subsubsection{Measurements}

We compare EMPIRE TSI to the records from the more recent monitoring missions that extend at least a decade (Fig. \ref{empssivscomposite}a) and the composites presently available (Fig. \ref{empssivscomposite}b). The various time series are normalized to the TIM record at the 2008 solar cycle minimum and smoothed with a 81-day boxcar filter to show up the long-term (solar cycle to decadal timescales) variability.

TSI measurements suffer long-term uncertainty, with the consequence that records start to diverge over the solar cycle (Fig. \ref{empssivscomposite}a) and composites, differing by which individual records are combined and how that is done, indicate conflicting decadal trends (Fig. \ref{empssivscomposite}b). The various records and composites are, however, sufficiently stable that they lie, at most times, well within 0.2 ${\rm Wm^{-2}}$ of one another in this comparison.

EMPIRE TSI is at about the same level at the 1986, 1996 and 2008 solar cycle minima (black, Fig. \ref{empssivscomposite}b), similar to what is indicated by the IRMB composite (green). As sunspot activity is relatively weak at solar cycle minima, the minimum-to-minimum trend in EMPIRE is effectively dictated by that in the Mg II index. Apart from the IUP Mg II index composite, there are three other Mg II index composites available, depicted in Fig. \ref{empssivscomposite}c. Similar to the situation with TSI, Mg II index measurements are afflicted by long-term uncertainty such that the various composites, differing by which individual records from the various UV SSI monitoring missions are combined and how that is done, show qualitatively different decadal trends. Considering this, the fact that EMPIRE exhibits little minimum-to-minimum variation cannot be taken as support of the similar trend indicated by the IRMB composite. Of course, this limitation does not only apply to EMPIRE but to proxy models based on the Mg II index in general. What we can reasonably conclude is that the long-term variability in EMPIRE TSI is consistent with observations to within the present limits of measurement stability, not differing from the records and composites examined here more than they do with one another (Figs. \ref{empssivscomposite}a and \ref{empssivscomposite}b).

TSI records are more consistent with one another at rotational timescales, where instrumental effects are relatively benign. As demonstrated in the example in Fig. \ref{empssivssim}e, observed TSI rotational variability is well-replicated by EMPIRE.

In Table \ref{tsicorrelation}, we summarize how EMPIRE compares with the TSI records and composites depicted in Figs. \ref{empssivscomposite}a and \ref{empssivscomposite}b in terms of the correlation and the RMS difference.

Of the ACRIM3, TIM and VIRGO records, the reconstruction agrees best with TIM ($R^2=0.91$), albeit only by a small margin more than with VIRGO ($R^2=0.88$). It is worth noting here that the parameters of the TSI model (Equation \ref{tsieqn}) are optimized to the TIM record (Sect. \ref{tsivar}), therefore biasing the reconstruction towards this record.

As for the composites, the EMPIRE reconstruction agrees best with PMOD ($R^2=0.86$), followed by IRMB ($R^2=0.81$) and ACRIM ($R^2=0.70$). This is within expectation. While the PMOD composite includes corrections to certain known instrumental issues in the original satellite records \citep{frohlich00,frohlich06}, the ACRIM and IRMB composites employ the original satellite records as they are. The IRMB composite is constructed by taking all available TSI records and taking the average where they overlap \citep{dewitte16}. This averaging dilutes the measurement noise and artefacts present in individual records, contributing to the better agreement with EMPIRE as compared to the ACRIM composite.

\subsubsection{Other models}
\label{tsicomparemodels}

In Fig. \ref{empssilongterm}a, we compare the EMPIRE (black) and NRLSSI2 (blue) reconstructions of TSI. Even though they are both given by the regression of the Mg II index and PSI to the TIM record (green), the amplitude of the solar cycle is systematically higher in EMPIRE. As evident from the figure, the amplitude of solar cycle 24 (2008 to the present) in the EMPIRE reconstruction is closer to what is indicated by the TIM record than NRLSSI2. This means the best fit of the Mg II index and PSI to the TIM record might be mis-estimated in the formulation of NRLSSI2. The discrepancy between the two reconstructions before 1976 is exacerbated by differences in how the PSI composite is generated.

The PSI is calculated from sunspot area measurements. Up to 1976, this is provided by the Royal Greenwich Observatory (RGO) and thereafter by SOON. RGO and SOON measurements are not equivalent to one another due to differences in the observation method, apparent in how the two compare to sunspot observations that overlap in time with both records. Various authors have sought to find the appropriate factor by which to multiply SOON sunspot areas to calibrate it to the RGO record \citep{fligge97,balmaceda09,hathaway10,foukal14}. In EMPIRE, we employ the PSI composite by \cite{balmaceda09}, who applied a calibration factor of 1.49. In NRLSSI2, the PSI composite is generated multiplying RGO sunspot areas by a factor of 0.8. This is based on the findings of \cite{fligge97}, who estimated the calibration factor to be between 1.15 and 1.25. Both models fit the Mg II index and PSI to the TIM record, which is in the period where the PSI is based on SOON measurements. Therefore, it is sunspot darkening in the RGO era that is affected by this correction. As demonstrated in Fig. \ref{empssilongterm}a, in EMPIRE, if we generate the PSI composite as done in NRLSSI2 instead, solar cycle amplitudes before 1976 become much closer to what is indicated by NRLSSI2.

\cite{fligge97} derived their estimate of the calibration factor, 1.15 to 1.25, by comparing the RGO and SOON records to sunspot area measurements from three other observatories. \cite{balmaceda09} came to the value of 1.49 by repeating this earlier analysis on an expanded data set, comparing data from more observatories and over longer periods of overlap. \cite{hathaway10} recovered a similar value of 1.48 by comparing the RGO and SOON sunspot area records to the international sunspot number. Given that the \cite{fligge97} study is effectively superseded by the \cite{balmaceda09} work and the calibration factor found by the latter is confirmed independently by \cite{hathaway10}, there is little basis for the step taken in NRLSSI2 to multiply RGO sunspot areas by a factor of 0.8 when generating the PSI composite. We note here that \cite{foukal14} argued that even though sunspot observations indicate a value of 1.4 to 1.5 for the calibration factor, it should be lowered to account for the sunspot size-dependence of the discrepancy between RGO and SOON measurements and sunspot intensity contrast. The author surmised that the appropriate value should be in the region of 1.2, but the adjustment to the calibration factor these considerations require, if at all, is really not known.

For each solar cycle in the period covered by the EMPIRE model, we calculated the spectral distribution of the change in solar irradiance between the minimum at the beginning of the cycle and the maximum. In EMPIRE, the distribution is broadly similar between the various cycles (Fig. \ref{sunspotareacalibrationfactor}a). If instead of employing the \cite{balmaceda09} PSI composite we generate the PSI composite as done in NRLSSI2, the distribution in the cycles before 1976 (red, Fig. \ref{sunspotareacalibrationfactor}b) become markedly different to those after (black), with much higher values in the UV and lower in the visible. It is unlikely that the physics of the Sun changed around 1976 in such a way as to produce such a drastic switch in the spectral distribution of solar irradiance variability. That is to say, this discrepancy between the cycles before and after 1976 is, in all likelihood, not solar in origin but a result of the calibration factor used by NRLSSI2 being inappropriate. The \cite{balmaceda09} PSI composite, and therefore the higher solar cycle amplitudes before 1976 in EMPIRE, are evidently more robust.


We excluded the NRLSSI1 reconstruction from this comparison as it is essentially similar to the NRLSSI2 reconstruction. The two time series do differ in terms of the secular trend after 1978 but that is only because NRLSSI1 employs the LASP Mg II index composite and NRLSSI2 the one by IUP.

The TSI models reported in the literature exhibit discrepant long-term trends due to differences in the modelling approach (i.e., proxy or semi-empirical), the solar observations employed and how secular variability is determined. The merits and limitations of the different steps taken by the various models is still a matter of debate and well beyond the scope of this paper. We refer the reader to the recent reviews by \cite{ermolli13}, \cite{solanki13} and \cite{yeo14b,yeo16}. Here, for information, we compare the EMPIRE reconstruction of TSI to that from SATIRE-S (Fig. \ref{empssilongterm}b). As SATIRE-S only goes back to 23 August 1974, the date of the first full-disc magnetogram suitable to the model available, we extend the time series further back in time with the reconstruction from the SATIRE-T model \citep{krivova10}. SATIRE-T is similar to SATIRE-S except the prevalence of solar surface magnetic features is inferred from the group sunspot number \citep{hoyt93} instead of full-disc magnetograms. As noted between existing models, EMPIRE and SATIRE exhibit discrepant long-term variability. The broad overall trend is, at least up to solar cycle 23 (1996 to 2008), rather similar even though they differ in terms of the shape and amplitude of the solar cycle. The divergence over solar cycle 24 is within the uncertainty of available observations (cf. Fig. \ref{empssivscomposite}).

In summary, the EMPIRE and NRLSSI2 models reconstruct TSI by essentially the same method but the amplitude of the solar cycle is underestimated in NRLSSI2 from shortcomings in how the best fit of the Mg II index and PSI to the TIM record and the PSI composite are determined. While the EMPIRE reconstruction is the more robust of the two, one must bear in mind that the discrepant long-term trend between the models reported in the literature, seen here in how EMPIRE and SATIRE compare to one another, is still a matter of debate.

\section{Summary}
\label{summary}

In this paper, we present a proxy model of TSI and SSI variability entitled EMPirical Irradiance REconstruction (EMPIRE). The reconstruction spans 115 to 170000 nm and from 14 February 1947 to the present at daily cadence.

EMPIRE UV SSI variability is determined from the regression of the rotational variability in the Mg II index and PSI to that in the SSI observations from the UARS and SORCE missions. This follows earlier proxy models such as MGNM, NRLSSI1 and NRLSSI2, except we applied orthogonal distance regression (ODR) instead of ordinary least squares (OLS). The UV SSI variability inferred from applying ODR is stronger than that from OLS. We demonstrated the difference to be due to biases in the OLS regression arising from the algorithm neglecting the uncertainty in the predictors (the Mg II index and PSI in this context). This is a well-established property of OLS, termed regression attenuation. ODR is designed to circumvent regression attenuation by taking the uncertainty in the predictors into account.

EMPIRE UV SSI variability, whether at solar cycle or rotational timescales, is well-supported by the available UV SSI records.

The solar cycle variability in EMPIRE UV SSI is stronger than in NRLSSI1 and NRLSSI2, which are established by the OLS analysis of UARS/SOLSTICE and SORCE SSI rotational variability, respectively. By reproducing the NRLSSI1 and NRLSSI2 process here, we showed that the discrepancy arose from an underestimation of UV SSI variability in NRLSSI1 and NRLSSI2 due to regression attenuation. Various studies have noted that the solar cycle variability in UV SSI in MGNM and NRLSSI1 is weaker than in other models such as that by \cite{morrill11} and SATIRE-S. Unlike existing proxy models, EMPIRE is consistent with both the \cite{morrill11} and SATIRE-S reconstructions. This strengthens the argument that UV variability is underestimated in earlier proxy models.

In the visible and IR, EMPIRE variability is determined from the calculated intensity of solar surface features following the approach of NRLSSI1. The reconstructed variability differs markedly from the NRLSSI1 and NRLSSI2 reconstructions. While EMPIRE and NRLSSI1 variability are derived by the same broad approach, the process in EMPIRE is arguably more robust from the choice of calculated intensities and how they are combined. As in the UV, NRLSSI2 visible and IR variability, up to 2400 nm, is determined from the OLS analysis of SORCE SSI rotational variability and so too is likely to suffer from regression attenuation.

As with reported models to date, EMPIRE does not reproduce the solar cycle variability in visible and IR SSI recorded by SIM. Multiple studies have concluded however, that SIM solar cycle variability is likely affected by unaccounted instrumental effects. EMPIRE does reproduce the rotational variability in SIM visible and IR SSI up to 1600 nm (longwards of 1600 nm, SIM measurements are dominated by noise).

EMPIRE TSI is given by the regression of the Mg II index and PSI to the TIM record. The long-term variability is consistent with the records from the more recent TSI monitoring missions and available composites to within the present limits of measurement stability. The amplitude of the solar cycle is, however, systematically higher than in NRLSSI2, where TSI is similarly given by the regression of the Mg II index and PSI to the TIM record. We demonstrated that this is due to a possible mis-estimation of the best fit of the Mg II index and PSI to the TIM record in the derivation of NRLSSI2 and this same model applying an inappropriate calibration to bring the sunspot areas from RGO and SOON, on which the PSI is based, to a common scale.

We demonstrated that the discrepant UV SSI solar cycle variability between earlier proxy models based on measured rotational variability and other models arose from regression attenuation in the former. EMPIRE, by circumventing regression attenuation, replicates UV SSI variability consistent with other models and observations. EMPIRE TSI and visible and IR SSI variability is also in agreement with observations. The EMPIRE reconstruction, providing daily TSI and UV to IR SSI since 1947, is of utility to climate modelling.

\appendix

\section{Uncertainty in the rotational variability in the Mg II index and PSI}
\label{efrotesrot}

We estimated the uncertainty in the rotational variability in the IUP Mg II index composite, $F^{\rm rot}$ by comparing it to the competing composite by LASP. After bringing the LASP composite to the absolute scale of the IUP composite by regression, we calculated the RMS difference in the rotational variability from 12 October 1991 on. {This is the period of interest since this is where we compared the IUP Mg II index composite and the PSI to measured UV SSI in the derivation of EMPIRE UV SSI variability (Sect. \ref{uvssivar}). Over this period, the LASP composite is composed of Mg II index measurements from a different set of instruments than the IUP composite, such that how the two composites differ is an indication of the uncertainty in these data.} We took the result, $5.7\times10^{-4}$ as the uncertainty in $F^{\rm rot}$. The standard deviation of $F^{\rm rot}$ over the same period is $2.0\times10^{-3}$.  The S/N, given by the ratio of signal variance to noise variance, is 12.5.

{The PSI, $S$ on a particular day is given by the sum of the intensity deficit of the prevailing sunspots, calculated from the observed area and distance from disc centre \citep{hudson82,frohlich94}. It is straightforward to show that the PSI is, to first order, proportional to the daily total sunspot area. The uncertainty in sunspot area measurements range from about $20\%$ for larger sunspots to $50\%$ for smaller sunspots \citep{sofia82}. Taking the relationship between the PSI and the daily total sunspot area, and the proportional uncertainty in sunspot area measurements into account, we assumed that the uncertainty in the PSI, $\epsilon_{S}$ is proportional to it such that $\epsilon_{S}=mS$, where $m$ is the constant of proportionality.}

{The \cite{balmaceda09} PSI composite is based on sunspot area measurements from RGO, SOON and the Pulkovo Observatory. The correlation, $R$ between the daily total sunspot area from the three observatories is in the region of 0.95 \citep[see Table 2 in][]{balmaceda09}. This deviation in the correlation from unity is of course from the uncertainty in sunspot area measurements. As the PSI is (to first order) proportional to the daily total sunspot area, the PSI based on sunspot area measurements from two different observers would have a similar correlation to one another as the daily total sunspot area from the two observers. With this in consideration, we fixed the value of $m$ by introducing scatter of standard deviation $mS$ to the PSI composite and finding the value of $m$ that brings the correlation between the result and the original time series closest to 0.95. We arrived at a final value of 0.26. This value is reasonable; given the proportional relationship between the PSI and the daily total sunspot area, $m$ must lie in the range of the proportional uncertainty in the observed area of individual sunspots (20\% to 50\%).}

We estimated the uncertainty in the rotational variability, $\epsilon_{S^{\rm rot}}$ from $\epsilon_{S}$ by propagation of errors. Over the UARS and SORCE missions, the average $\epsilon_{S^{\rm rot}}$ is $9.1\times10^{-3}$ and the standard deviation of $S^{\rm rot}$ is $3.2\times10^{-2}$. This corresponds to a S/N of 12.4.

\section{Denoising the rotational variability in observed SSI by a non-local means filter approach}
\label{nlmf}

The non-local means filter or NLMF \citep{baudes05} is an algorithm for image denoising. The signal at a given point in an image is replaced by the weighted sum of all the points in the image. The weight is calculated comparing the neighbourhood of the given point and the neighbourhood of each point in the image such that the more similar they are, the greater the weight. The assumption is that in any natural image, there is redundancy such that there is no part that is truly unique, allowing the suppression of noise by matching each point to other points with similar neighbourhoods and taking the similarity-weighted average. We denoised measured SSI rotational variability, $I^{\rm rot}_{\rm obs}$ in the UARS and SORCE records by this approach, except here the weighting is given by the similarity in the rotational variability in solar activity instead.

For a given SSI record and wavelength band, the denoised $I^{\rm rot}_{\rm obs}$ on the $j$-th day in the time series, $I^{\rm rot}_{\rm dns}\left(t_{j}\right)$ is given by
\begin{equation}
	I^{\rm rot}_{\rm dns}\left(t_{j}\right)=\sum^{n}_{i=1}w_{j}\left(t_{i}\right)I^{\rm rot}_{\rm obs}\left(t_{i}\right),
\end{equation}
where $w_{j}\left(t_{i}\right)$ denotes the weight assigned to the $i$-th day. The weights are defined as
\begin{equation}
	w_{j}\left(t_{i}\right)=\exp\left(-\frac{\delta_{j}\left(t_{i}\right)}{\tau}\right)/\left[\sum^{n}_{i=1}\exp\left(-\frac{\delta_{j}\left(t_{i}\right)}{\tau}\right)\right],
\label{weights}
\end{equation}
where $\delta_{j}\left(t_{i}\right)$ is a measure of the similarity in the rotational variability in solar activity between the $j$-th and $i$-th days. The parameter $\tau$ controls the decay of $w_{j}$ with $\delta_{j}\left(t_{i}\right)$ while the normalization ensures the sum of weights equate to unity. Let $J^{\rm rot}_{\rm obs}$ denote the detrended integrated SSI over the wavelength range of the record and $\sigma_x$ the standard deviation of variable $x$. The similarity measure, $\delta_{j}\left(t_{i}\right)$ is given by
\begin{eqnarray}
	\delta_{j}\left(t_{i}\right)&=&\left[\frac{F^{\rm rot}\left(t_{j}\right)-F^{\rm rot}\left(t_{i}\right)}{\sigma_{F^{\rm rot}}}\right]^2+\nonumber\\
	&&\left[\frac{S^{\rm rot}\left(t_{j}\right)-S^{\rm rot}\left(t_{i}\right)}{\sigma_{S^{\rm rot}}}\right]^2+\nonumber\\
	&&\left[\frac{J^{\rm rot}_{\rm obs}\left(t_{j}\right)-J^{\rm rot}_{\rm obs}\left(t_{i}\right)}{\sigma_{J^{\rm rot}_{\rm obs}}}\right]^2;
\label{similarity}
\end{eqnarray}
the square of the variance weighted Euclidean distance between the $j$-th and $i$-th days in $F^{\rm rot}$-$S^{\rm rot}$-$J^{\rm rot}_{\rm obs}$-space. From Equations \ref{weights} and \ref{similarity}, $\delta_{j}\left(t_{i}\right)$ is lower and $w_{j}\left(t_{i}\right)$ consequently higher where the rotational variability in solar activity is similar to that in the $j$-th day as indicated by $F^{\rm rot}$, $S^{\rm rot}$ and $J^{\rm rot}_{\rm obs}$. The greatest weight is accorded to the original signal, as $\delta_{j}\left(t_{i}\right)=0$ when $i=j$. Let us denote the $\delta_{j}\left(t_{i}\right)$ time series excluding the $i=j$ instance as $\delta_{j}\left(t_{i,i\neq{}j}\right)$.

Calculating $\delta_{j}\left(t_{i}\right)$ for the UARS and SORCE records, we noted that for most days, the minimum value of $\delta_{j}\left(t_{i,i\neq{}j}\right)$ is in the order of $10^{-3}$ to $10^{-1}$. Taking into account that this is the range of $\delta_{j}\left(t_{i}\right)$ between the days most similar to one another in terms of the rotational variability in solar activity, we set the parameter $\tau$ in Equation \ref{weights} at 0.1. In defining $w_{j}\left(t_{i}\right)$ as an exponential function of $-\frac{\delta_{j}\left(t_{i}\right)}{\tau}$ and setting $\tau$ at this level, the weighting is heavily skewed towards the most similar days. This conservative measure limits possible artefacts from giving undue weight to dissimilar days.

There are days with unusual rotational variability in solar activity over the lifetime of UARS and SORCE. This is indicated by $\delta_{j}\left(t_{i,i\neq{}j}\right)$ of a particular day well exceeding the order of $10^{-1}$ across the period of the given SSI record. This does not result in the recreated signal on such days being erroneously dominated by the signal from days that are dissimilar in terms of the rotational variability in solar activity. Let us take 29th October 2003, from the period of exceptional solar activity around late-October to early-November 2003, in the UARS/SUSIM record as an example. For this day in this record, $\delta_{j}\left(t_{i,i\neq{}j}\right)$ ranged from about 16 to 960. Due to the definition of $w_i$ and our choice of $\tau$, the total weight assigned to all the days apart from 29th October 2003 itself is merely $6\times10^{-73}$. It follows that the weighting accorded to this day itself is effectively unity and the original signal is retained in the recreated time series.

It is worth emphasizing here that including $F^{\rm rot}$, $S^{\rm rot}$ and $J^{\rm rot}_{\rm obs}$ in the definition of $\delta_{j}\left(t_{i}\right)$ (Equation \ref{similarity}) does not impose the variability of these time series on $I^{\rm rot}_{\rm dns}\left(t_{j}\right)$. It is not possible to enforce a trend on $I^{\rm rot}_{\rm dns}\left(t_{j}\right)$ by designing $\delta_{j}\left(t_{i}\right)$. If $\delta_{j}\left(t_{i}\right)$ is not an accurate reflection of similarity in the underlying true signal in $I^{\rm rot}_{\rm obs}$, we are effectively recreating the signal on each day with signals from days with varying true signals. The result is a $I^{\rm rot}_{\rm dns}\left(t_{j}\right)$ time series where variability is smoothed out, the severity depending on how erroneous the designed $\delta_{j}\left(t_{i}\right)$ is. In such a scenario the variability of $I^{\rm rot}_{\rm dns}\left(t_{j}\right)$ will follow neither the underlying true signal in $I^{\rm rot}_{\rm obs}$ nor the designed $\delta_{j}\left(t_{i}\right)$. This sensitivity to $\delta_{j}\left(t_{i}\right)$ does however also mean that some variability is inevitably lost in the NLMF process due to the uncertainty in the quantities used to calculate $\delta_{j}\left(t_{i}\right)$. As such, the variability in the recreated time series should be taken as the lower limit of the true variability.

\section*{Acknowledgments}

We thank Micha Sch\"oll and Odele Coddington for the fruitful discussions. This work is supported by the German Federal Ministry of Education and Research under project 01LG1209A and in part by the Ministry of Education of Korea through the BK21 plus program of the National Research Foundation.

The ACRIM TSI composite and the ACRIMSAT/ACRIM3 TSI record are available at www.acrim.com, the Aura/OMI and SBUV SSI records at sbuv2.gsfc.nasa.gov, the F10.7 record at www.spaceweather.ca, the international sunspot number at sidc.oma.be, the IRMB TSI composite at gerb.oma.be/steven/RMIB\_TSI\_composite/, the IUP Mg II index composite at www.iup.uni-bremen.de/gome/gomemgii.html, the NOAA Mg II index composite at ftp.ngdc.noaa.gov/STP/SOLAR\_DATA/SOLAR\_UV/, the NRLSSI2 reconstruction at data.ncdc.noaa.gov/cdr/solar-irradiance/, the PMOD TSI composite and the SoHO/VIRGO TSI record at www.pmodwrc.ch, the SET Mg II index composite at www.spacewx.com/About\_MgII.html and the UARS/SUSIM SSI record at wwwsolar.nrl.navy.mil/uars/. The LASP Mg II index composite, the NRLSSI1 reconstruction, the WHI reference spectra and the SME, UARS/SOLSTICE, SORCE and TIMED/SEE SSI records are hosted on the LISIRD data centre (lasp.colorado.edu/lisird/). The \cite{morrill11} reconstruction is available on request (jeff.morrill@nrl.navy.mil). The EMPIRE, SATIRE-S and SATIRE-T reconstructions and the \cite{balmaceda09} PSI composite can be downloaded from www2.mps.mpg.de/projects/sun-climate/data.html.


\end{article}

\begin{sidewaystable}
\caption{Description of the SSI records employed in the derivation of EMPIRE UV SSI variability (see Sect. \ref{uvssivar}).}
\label{ssirecords}
\centering
\begin{tabular}{lccccc} 
 & & Wavelength & Wavelength & \\
SSI record (version) & Period [year.month.day] & range [nm] & resolution [nm]& Reference(s) \\
\hline
UARS/SUSIM (22) & 1991.10.12 to 2005.07.31 & 115 to 411 & 1 & \cite{brueckner93,floyd03} \\
UARS/SOLSTICE (18) & 1991.10.17 to 2001.09.24 & 119 to 420 & 1 & \cite{rottman01} \\
SORCE/SIM (22) & 2003.04.14 to 2015.05.02 & 240 to 2416 & 1 to 34 & \cite{harder05a,harder05b} \\
SORCE/SOLSTICE (15) & 2003.05.14 to 2015.10.31 & 115 to 310 & 1 & \cite{mcclintock05,snow05a} \\
\hline          
\end{tabular}
\end{sidewaystable}

\begin{sidewaystable}
\caption{Description of the UV SSI reconstructions depicted in Figs. \ref{empssisingleestimates}, \ref{empssiodrvsols} and \ref{empssivsnrlssi}.}
\label{reconstructiontable}
\centering
\begin{tabularx}{\textwidth}{lc>{\hsize=\hsize}X}
Index & Reconstruction & Description \\
\hline
1 & Blue solid (Figs. \ref{empssisingleestimates} and \ref{empssiodrvsols}a) & For each of the UARS and SORCE records, the OLS regression of $F^{\rm rot}$ and $S^{\rm rot}$ to $I^{\rm rot}_{\rm obs}\left(\lambda\right)$ (Sect. \ref{uvssivar}). \\
2 & Red solid (Figs. \ref{empssisingleestimates} and \ref{empssiodrvsols}a) & For each of the UARS and SORCE records, the ODR regression of $F^{\rm rot}$ and $S^{\rm rot}$ to $I^{\rm rot}_{\rm obs}\left(\lambda\right)$ (Sect. \ref{uvssivar}). \\
3 & Black (Figs. \ref{empssisingleestimates}, \ref{empssiodrvsols}b and \ref{empssivsnrlssi}) & EMPIRE reconstruction; mean of the results from the UARS and SORCE records in item 2 (Sect. \ref{uvssivar}). \\
4 & Blue dash (Fig. \ref{empssiodrvsols}a) & OLS regression of $F^{\rm rot}$ and $S^{\rm rot}$ to denoised UARS/SUSIM $I^{\rm rot}_{\rm obs}\left(\lambda\right)$ (Sect. \ref{cleaniobsrot}). \\
5 & Red dash (Fig. \ref{empssiodrvsols}a)  & ODR regression of $F^{\rm rot}$ and $S^{\rm rot}$ to denoised UARS/SUSIM $I^{\rm rot}_{\rm obs}\left(\lambda\right)$ (Sect. \ref{cleaniobsrot}). \\
6 & Blue dot (Fig. \ref{empssiodrvsols}b) & OLS regression of the $F^{\rm rot}$ and $S^{\rm rot}$ mean transit profiles to the $I^{\rm rot}_{\rm obs}\left(\lambda\right)$ mean transit profiles (Sect. \ref{cleanfrotsrot}). \\
7 & Green solid (Fig. \ref{empssivsnrlssi}a) & NRLSSI1 reconstruction \citep{lean97,lean00a}; OLS regression of $F^{\rm rot}$ and $S^{\rm rot}$ to UARS/SOLSTICE $I^{\rm rot}_{\rm obs}\left(\lambda\right)$ (Sect. \ref{nrlssi}). \\
8 & Violet solid (Fig. \ref{empssivsnrlssi}a) & NRLSSI2 reconstruction \citep{coddington16}; OLS regression of $F^{\rm rot}$ and $S^{\rm rot}$ to SORCE/SOLSTICE and SORCE/SIM $I^{\rm rot}_{\rm obs}\left(\lambda\right)$ (Sect. \ref{nrlssi}). \\
9 & Green dash (Fig. \ref{empssivsnrlssi}a) & Reproduction of the NRLSSI1 reconstruction (Sect. \ref{nrlssi}). \\
10 & Violet dash (Fig. \ref{empssivsnrlssi}a) & Reproduction of the NRLSSI2 reconstruction (Sect. \ref{nrlssi}). \\
11 & Violet dot (Fig. \ref{empssivsnrlssi}a) & Same as item 11, except excluding the correction to NRLSSI2 introduced by \citealt{coddington16} (Sect. \ref{nrlssi}). \\
12 & Orange (Fig. \ref{empssivsnrlssi}b) & SATIRE-S reconstruction \citep{yeo14a}; non-proxy reconstruction based on the calculated intensity spectra of solar surface features and maps of facular and sunspot surface coverage derived from full-disc solar observations (Sect. \ref{othermodels}).\\
13 & Cyan solid (Fig. \ref{empssivsnrlssi}b) & Mea11 reconstruction \citep{morrill11}; OLS regression of $F$ to UARS/SUSIM $I_{\rm obs}$ (Sect. \ref{othermodels}).\\
14 & Cyan dash (Fig. \ref{empssivsnrlssi}b) & OLS regression of $F$ and $S$ to UARS/SUSIM $I_{\rm obs}$ (Sect. \ref{othermodels}).\\
\hline          
\end{tabularx}
\end{sidewaystable}

\begin{table}
\caption{The 1-$\sigma$ uncertainty in the change between the 1996 solar cycle minimum and the 2000 maximum in TSI and in the integrated SSI over the specified wavelength ranges in the EMPIRE reconstruction (see Sect. \ref{error}).}
\label{uncertainty}
\centering
\begin{tabular}{lcc}
Description & Wavelength range [nm] & Uncertainty [\%] \\
\hline
TSI & & 1.1 \\
Lyman-$\alpha$ line & 121 to 122 & 0.5 \\         
Schuman-Runge continuum & 130 to 175 & 0.7 \\
Schuman-Runge bands & 175 to 200 & 0.7 \\
Herzburg continuum & 200 to 242 & 1.1 \\
Hartley bands & 200 to 300 & 2.5 \\
Higgins bands & 300 to 360 & 5.9 \\
H$_{2}$O and CO$_{2}$ bands & 700 to 5000 & 5.2 \\
\hline 
\end{tabular}
\end{table}

\begin{figure}
\includegraphics[width=\textwidth]{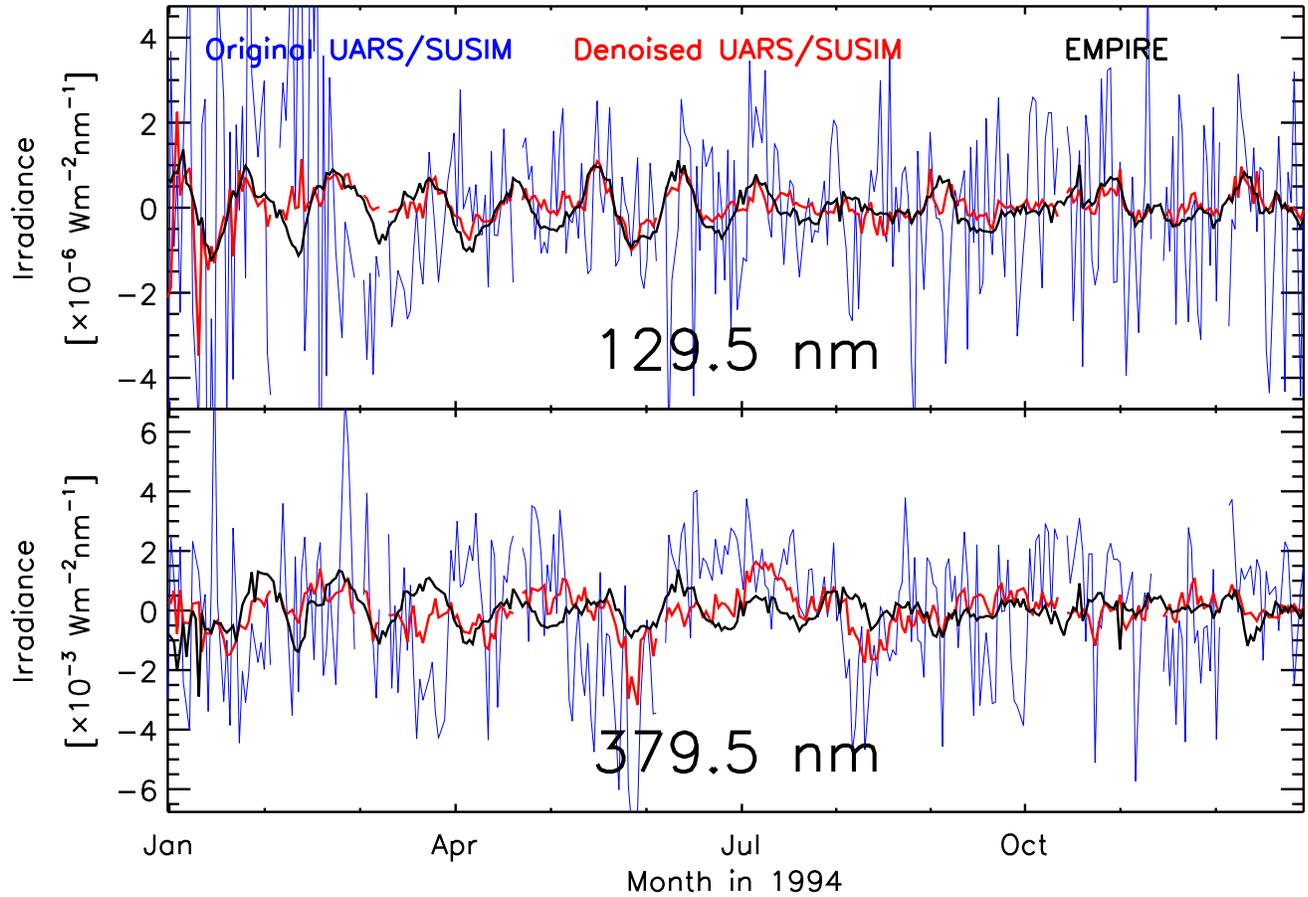}
\caption{Rotational variability in SSI at 129.5 nm (top) and 379.5 nm (bottom) over 1994. The blue and red lines follow the values from the SUSIM record before and after denoising (Sect. \ref{cleaniobsrot}), and the black lines the EMPIRE reconstruction. Here and in the rest of this paper, the gaps in rotational variability time series plots correspond to periods of three days and longer for which there are no data.}
\label{empssinlmf}
\end{figure}

\begin{figure}
\includegraphics[width=.7\textwidth]{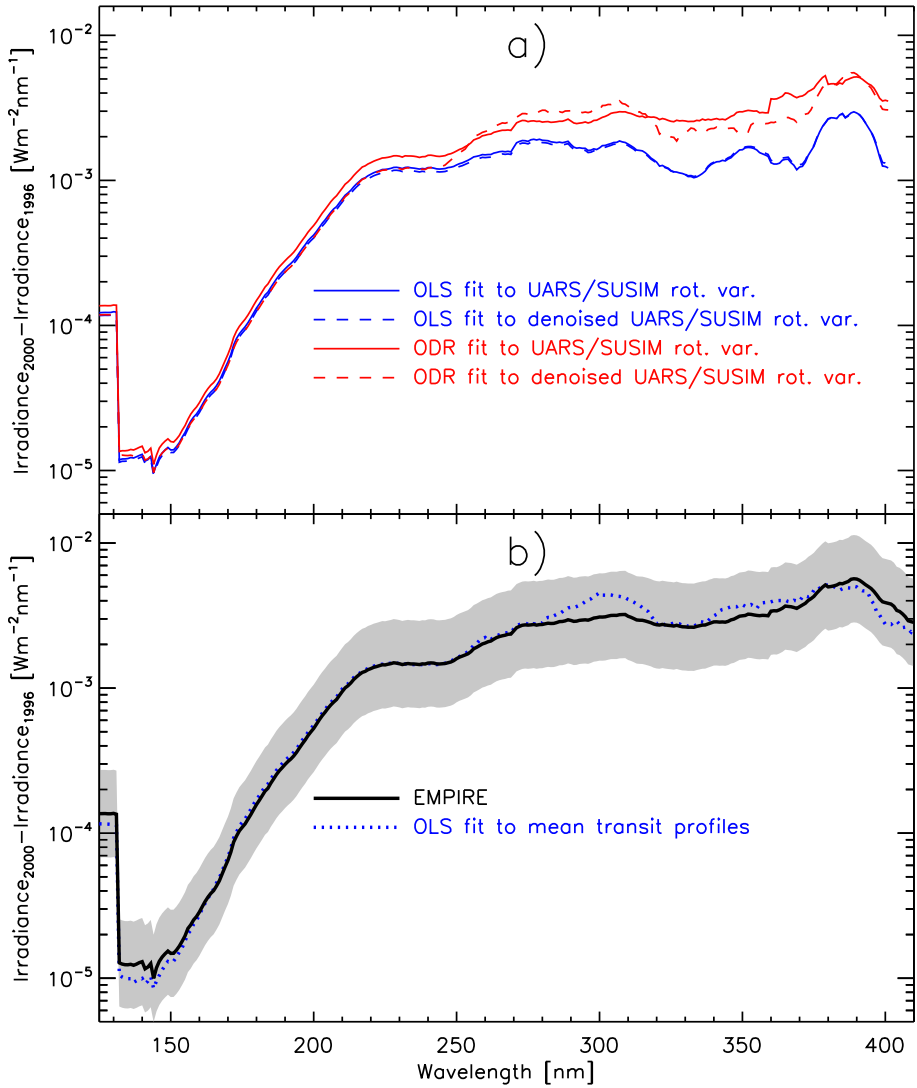}
\caption{The change in UV SSI over the ascending phase of solar cycle 23, as in Fig. \ref{empssisingleestimates}. a) The values determined from the OLS (blue) and ODR (red) analyses of SUSIM SSI rotational variability. The solid lines denote the results based on the original data set (taken from Fig. \ref{empssisingleestimates}) and the dashed lines that after denoising (Sect. \ref{cleaniobsrot}). b) The blue dotted line follows the variation inferred from the OLS analysis of the mean transit profiles (Sect. \ref{cleanfrotsrot}). The black line and shaded area have the same meaning as in Fig. \ref{empssisingleestimates}.}
\label{empssiodrvsols}
\end{figure}

\begin{figure}
\includegraphics[width=.6\textwidth]{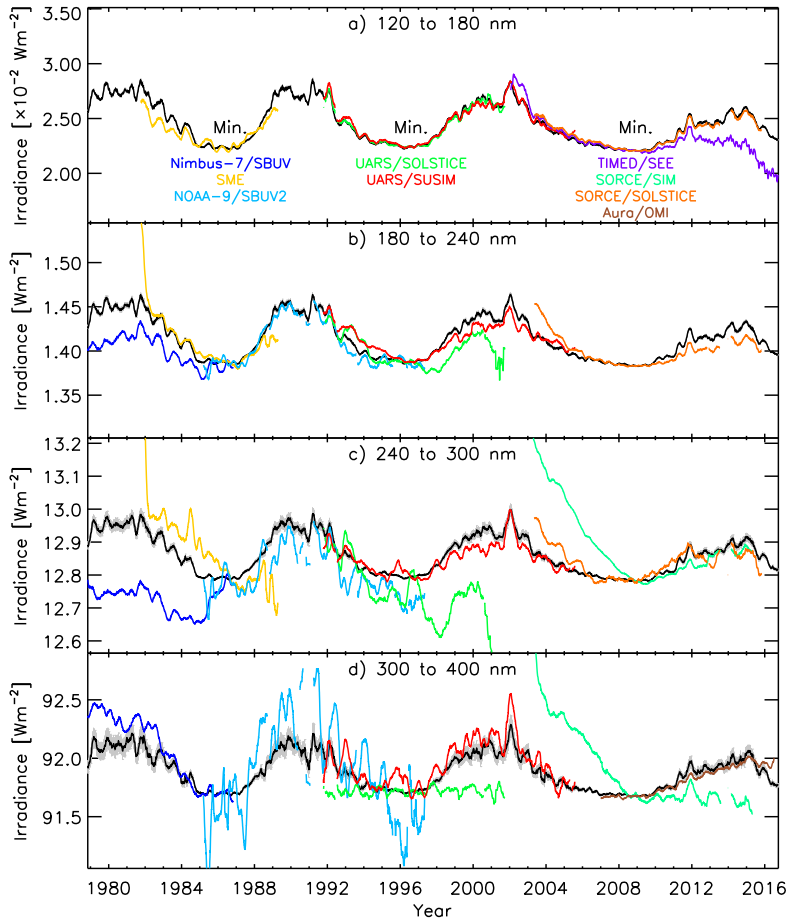}
\caption{Integrated SSI over the annotated wavelength intervals, smoothed with a 81-day boxcar filter, in EMPIRE (black) and in satellite records (colour). The shaded regions correspond to the 3-$\sigma$ bound of the EMPIRE reconstruction, largely hidden in the lower wavelength intervals where the uncertainty is minute. The gaps correspond to periods longer than 27 days without observations. On top of the UARS and SORCE records, described in Table \ref{ssirecords}, we depict the SSI records from Nimbus-7/SBUV \citep{deland01}, SME \citep{rottman88}, NOAA-9/SBUV2 \citep{deland04}, TIMED/SEE \citep[version 11,][]{woods05} and Aura/OMI \citep[20 July 2016 revision,][]{marchenko14,marchenko16}. The time series from the various records are normalized to EMPIRE at solar cycle minima, annotated in the top panel (1986 for the SBUV and SME records, 1996 for the UARS records and 2008 for the SEE, SORCE and OMI records).}
\label{empssivsuvssi}
\end{figure}

\begin{figure}
\includegraphics[width=\textwidth]{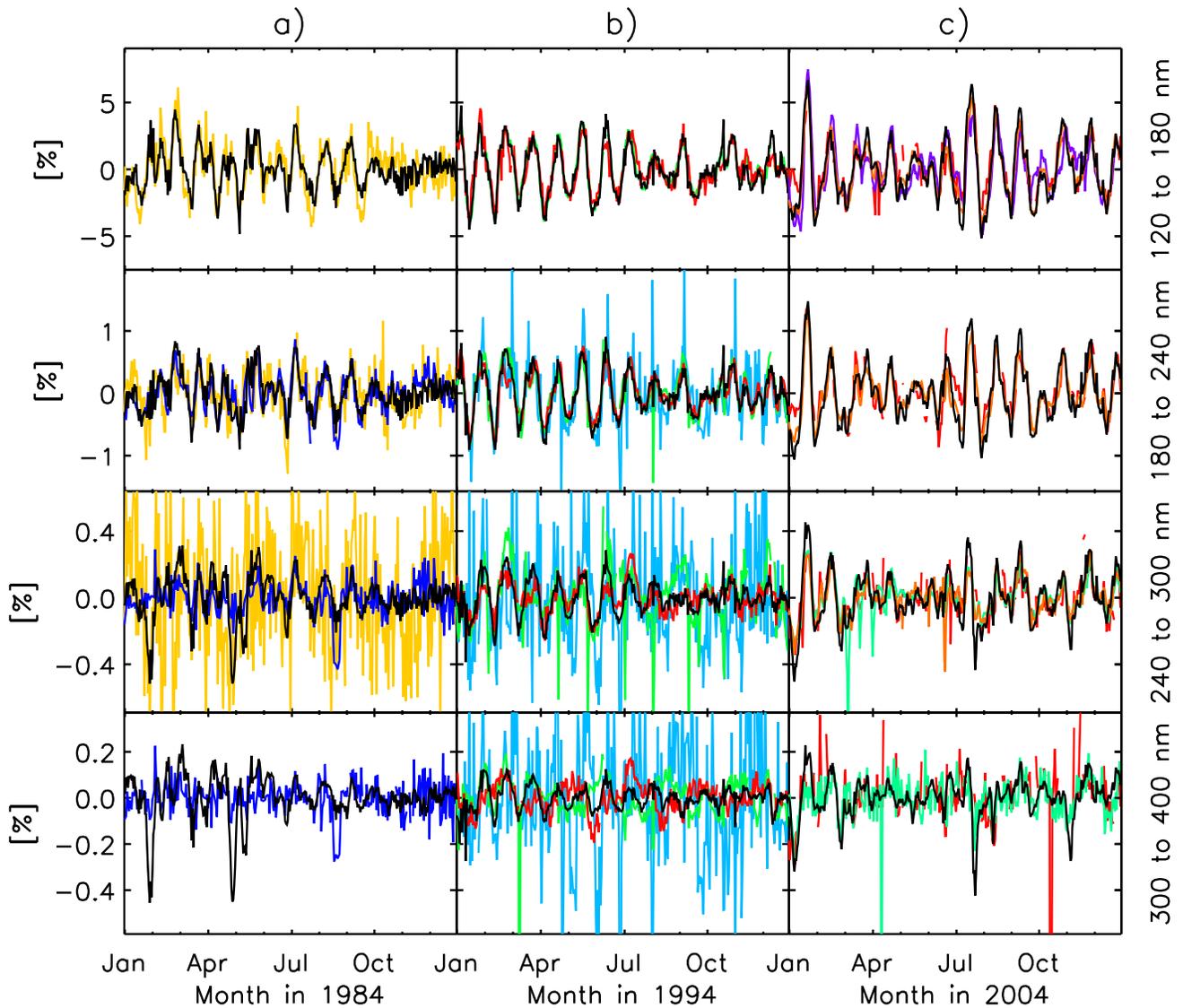}
\caption{Rotational variability in the integrated UV SSI over the annotated wavelength intervals (right-hand axis), in EMPIRE (black) and in satellite records (colour), over a) 1984, b) 1994 and c) 2004. The wavelength intervals and colour-coding follow Fig. \ref{empssivsuvssi}. The proportional difference between each integrated SSI time series and the corresponding 81-day moving average, highlighting the rotational variability relative to the overall level, is depicted.}
\label{empssivsuvssirot}
\end{figure}

\begin{figure}
\includegraphics[width=\textwidth]{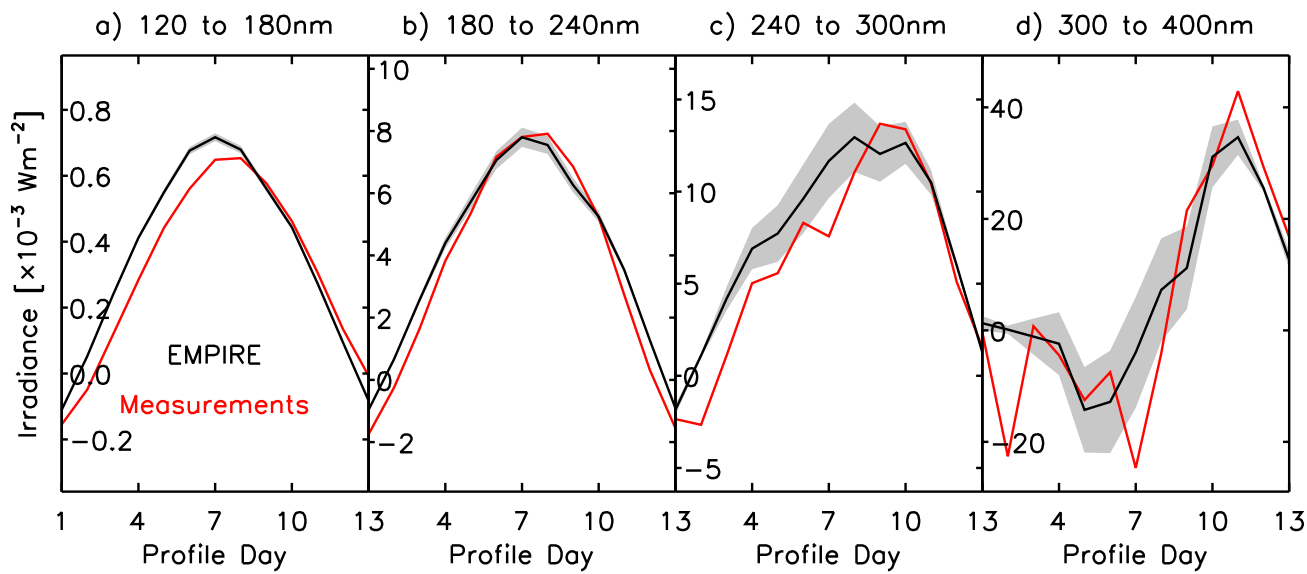}
\caption{Mean transit profile of the rotational variability in the integrated SSI over the annotated wavelength intervals (Sect. \ref{uvcompare}). The red and black profiles correspond to the mean transit profiles from the satellite records examined in Figs. \ref{empssivsuvssi} and \ref{empssivsuvssirot} (excluding Nimbus-7/SBUV, SME and Aura/OMI) and the corresponding EMPIRE reconstruction, respectively. The shaded regions enclose the 3-$\sigma$ bound of the EMPIRE reconstruction, barely visible in the lower wavelength intervals where the uncertainty is minute.}
\label{empssivsmeanprofile}
\end{figure}

\begin{figure}
\includegraphics[width=.7\textwidth]{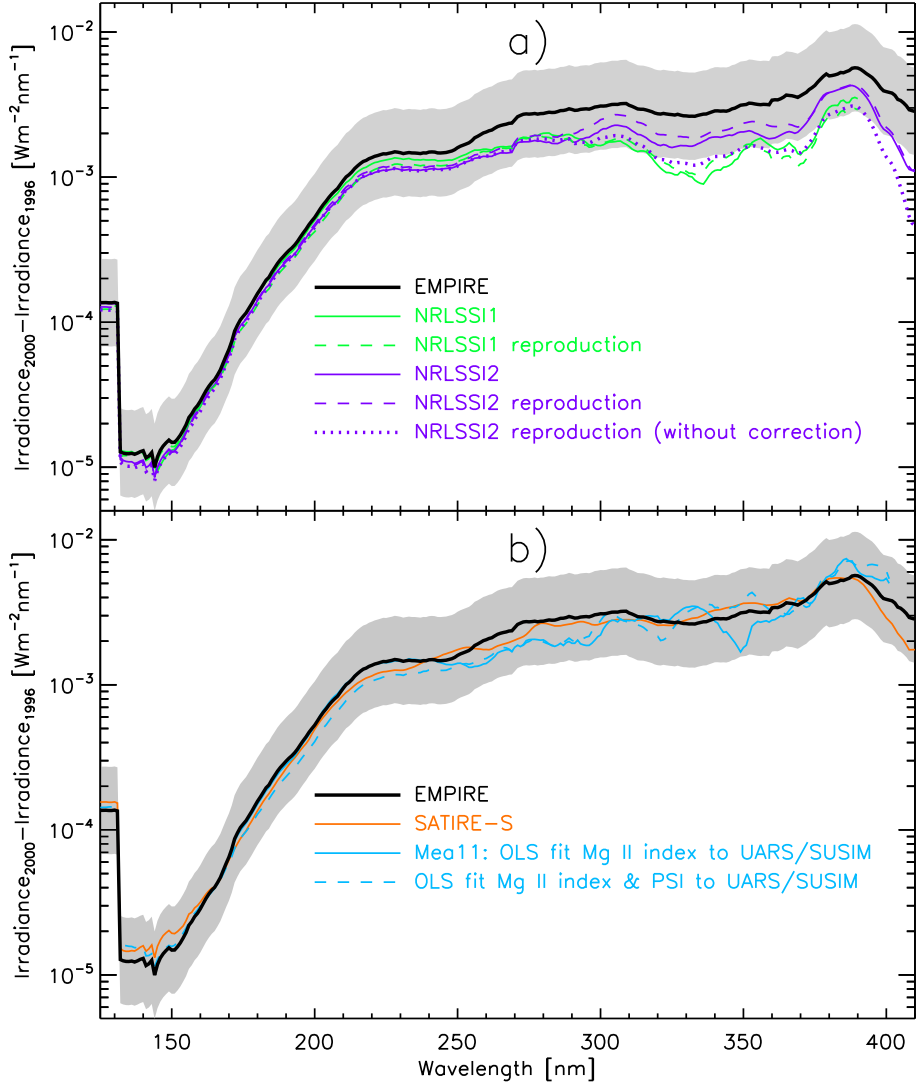}
\caption{Variation in UV SSI over the ascending phase of solar cycle 23 in EMPIRE (black) and the corresponding $50\%$ and $200\%$ bounds (shared area), taken from Fig. \ref{empssisingleestimates}. a) The NRLSSI1 (green solid line) and NRLSSI2 reconstructions (violet solid line), and the values from reproducing these models in this study (dashed lines). The violet dotted line denotes the NRLSSI2 reproduction without the correction introduced to this model by \cite{coddington16}. See Sect. \ref{nrlssi} for details. b) The orange line corresponds to the SATIRE-S reconstruction and the cyan solid line to Mea11, given by the regression of the Mg II index to SUSIM SSI. The values from fitting both the Mg II index and PSI to the SUSIM record is also depicted (cyan dashed line).}
\label{empssivsnrlssi}
\end{figure}

\begin{figure}
\includegraphics[width=.9\textwidth]{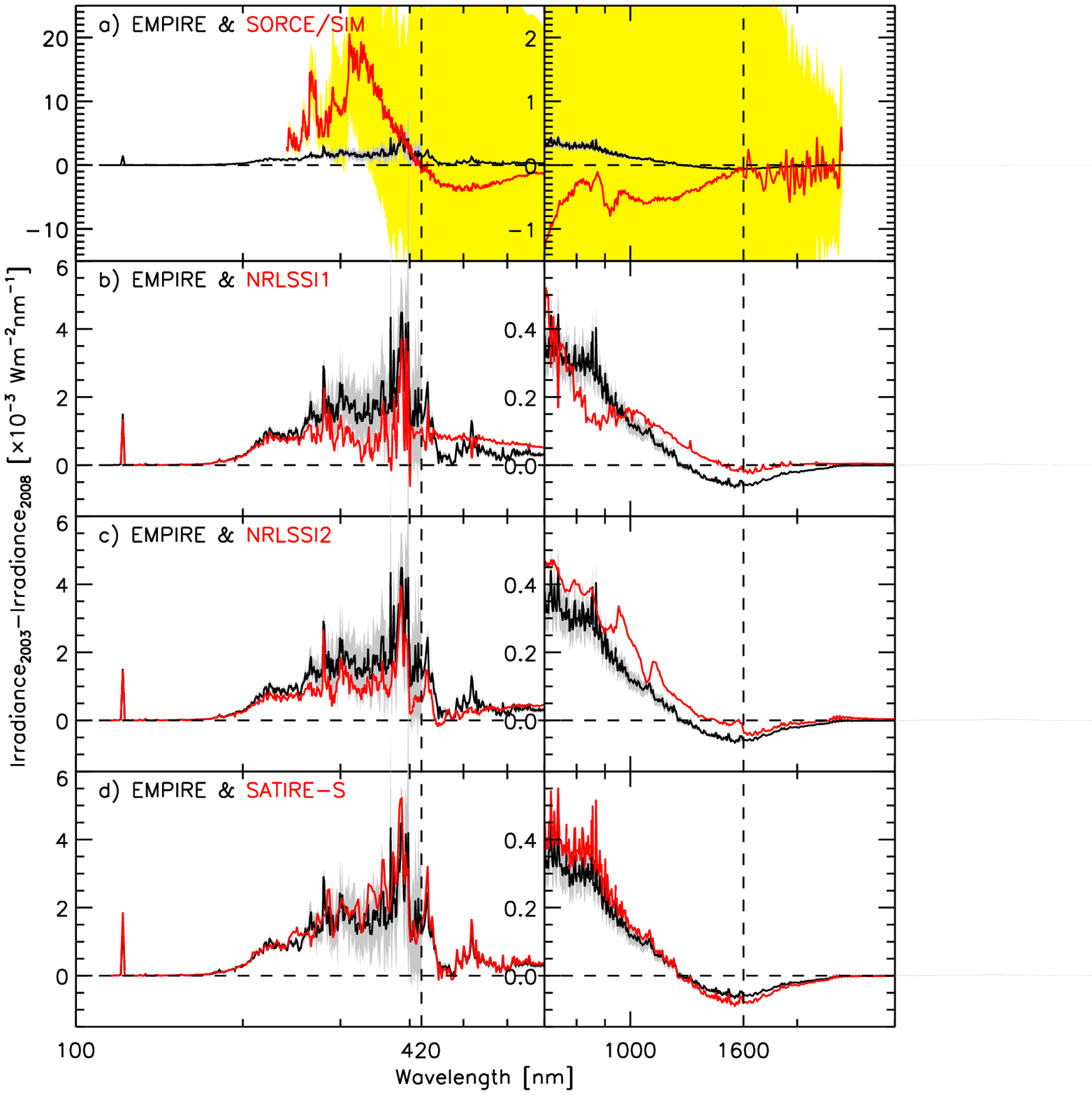}
\caption{The change in SSI between 2003 and the 2008 solar cycle minimum in EMPIRE (black), compared with the a) SIM, b) NRLSSI1, c) NRLSSI2 and d) SATIRE-S data sets (red). We depict the difference between the average spectrum from June 2003 to August 2003 and the average spectrum from November 2008 to January 2009. The horizontal dashed lines denote the zero level and the vertical dashed lines 420 and 1600 nm, drawn to aid the discussion (Sect. \ref{visircompare}). The SSI difference between 700 and 3000 nm is magnified by a factor of ten for visibility. The grey and yellow shaded regions correspond to the 3-$\sigma$ bound of EMPIRE and SIM, respectively.}
\label{empssispectraldelta}
\end{figure}

\begin{figure}
\includegraphics[width=\textwidth]{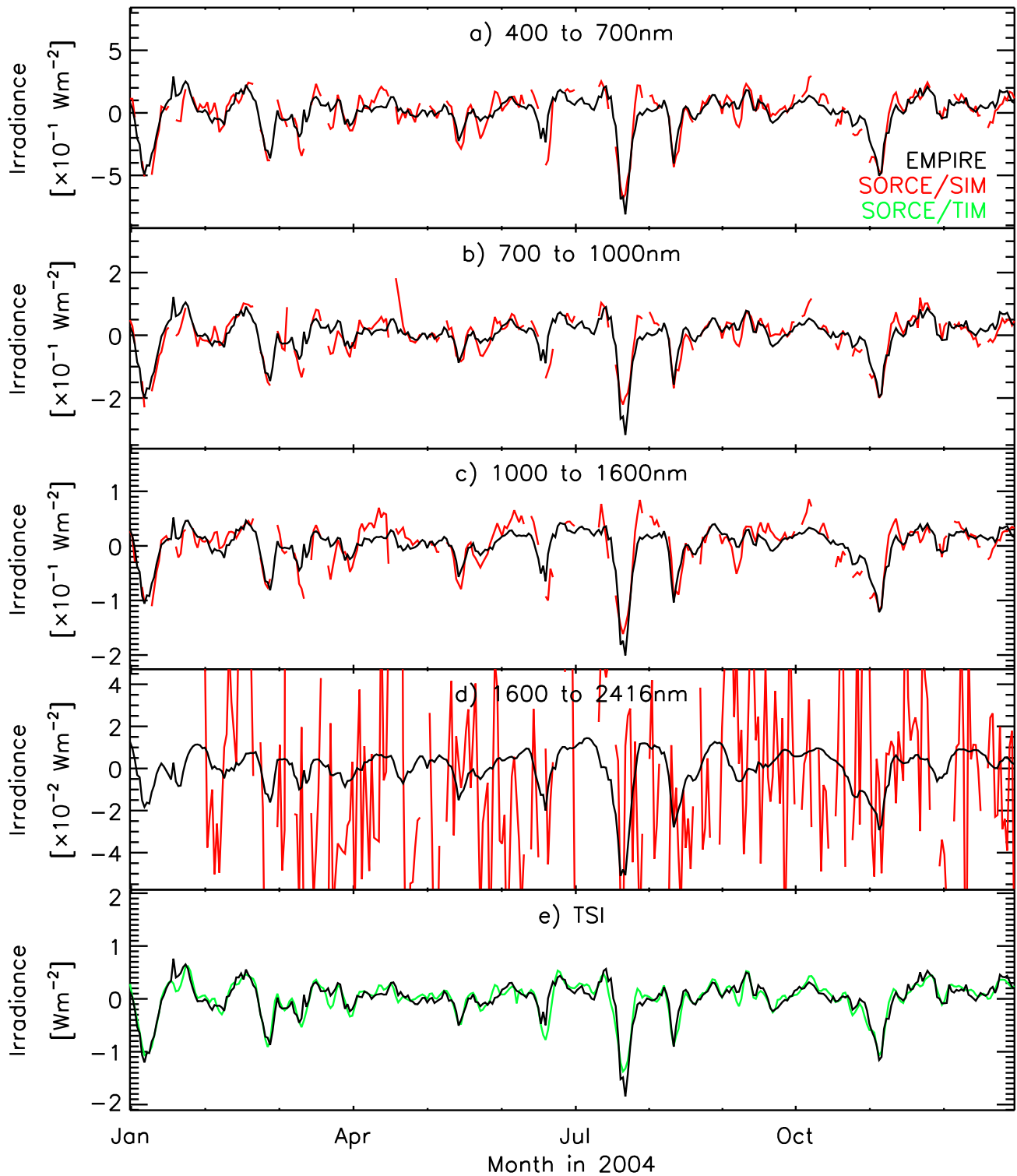}
\caption{Rotational variability in SSI integrated over the annotated wavelength intervals and in TSI over 2004. The black lines correspond to the EMPIRE reconstruction and the red lines to SORCE observations (SIM SSI and TIM TSI).}
\label{empssivssim}
\end{figure}

\begin{figure}
\includegraphics[width=.9\textwidth]{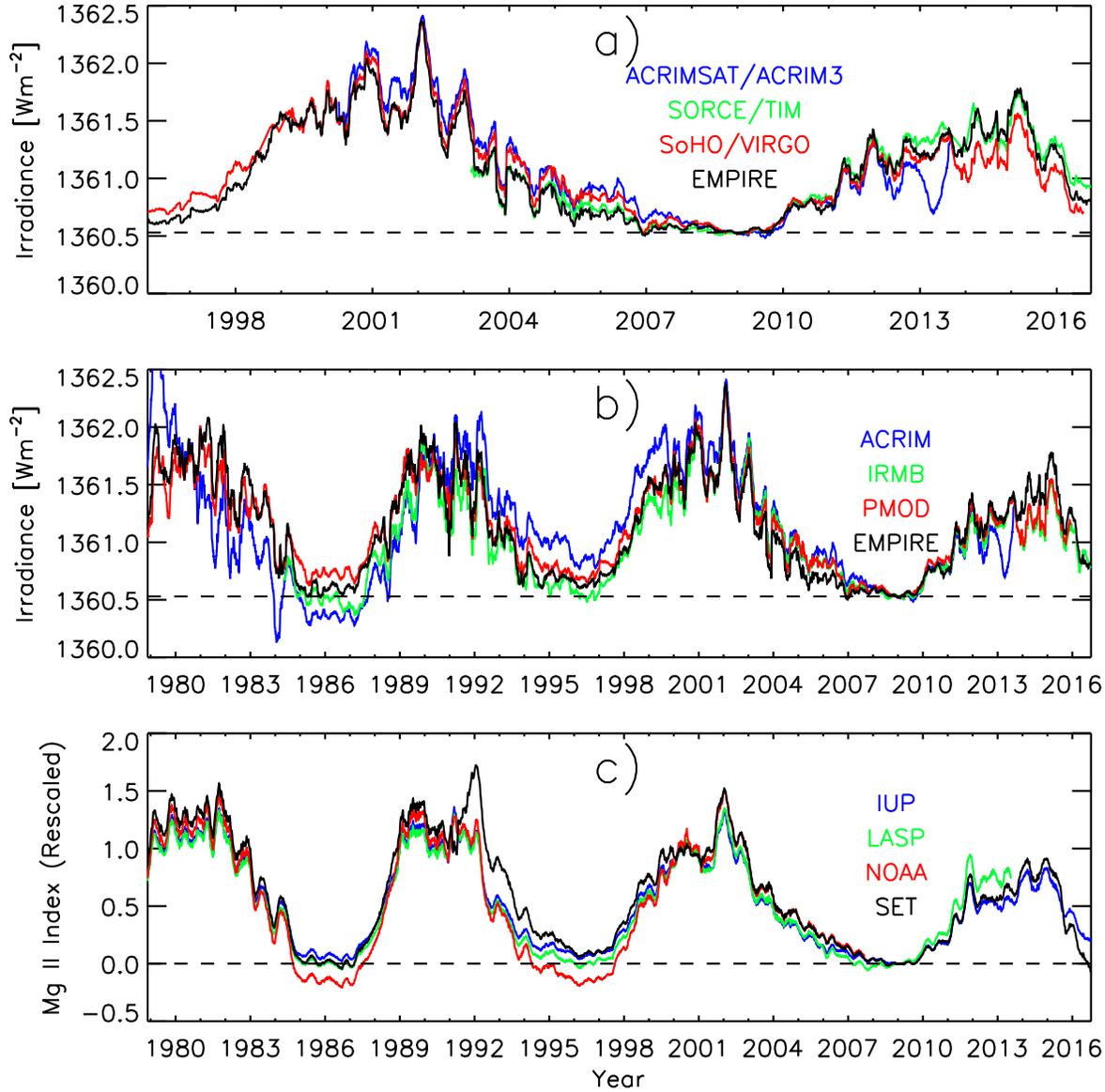}
\caption{The EMPIRE reconstruction of TSI, overplotted on a) the records from the more recent TSI monitoring missions and b) the available composites, described in Table \ref{tsicorrelation}. The various time series are normalized to the TIM record at the 2008 solar cycle minimum and smoothed with a 81-day boxcar filter. c) The Mg II index composites by IUP, LASP \citep{snow05b}, NOAA \citep{viereck04} and Space Environment Technologies (SET), stretched across null and unity at the 2008 minimum and 2000 maximum levels. Again, the 81-day moving average is depicted. The dashed lines follow the 2008 minimum level.}
\label{empssivscomposite}
\end{figure}

\begin{sidewaystable}
\caption{Description of the instrument and composite records of TSI featured in Fig. \ref{empssivscomposite}, and how they compare to EMPIRE in terms of the correlation, as given by $R$ and $R^2$, and the RMS difference. We compared the EMPIRE reconstruction to the ACRIM3, TIM and VIRGO records over the days where measurements are available from all three instruments, and similarly for the comparison with the ACRIM, IRMB and PMOD composites.}
\label{tsicorrelation}
\centering
\begin{tabular}{lccccc}
 & & & & & RMS difference \\
TSI record (version) & Period [year.month.day] & Reference(s) & $R$ & $R^2$ & [${\rm Wm^{-2}}$] \\
\hline
ACRIMSAT/ACRIM3 (11/13) & 2000.04.06 to 2013.09.17 & \cite{willson03} & 0.85 & 0.72 & 0.21 \\
SORCE/TIM (17, dated 2016.11.21) & 2003.02.25 to 2016.11.13 & \cite{kopp05c} & 0.95 & 0.91 & 0.12  \\
SoHO/VIRGO (6\_005\_1608) & 1996.01.28 to 2016.08.03 & \cite{frohlich95,frohlich97} & 0.94 & 0.88 & 0.14 \\
\hline
ACRIM composite (11/13) & 1978.11.17 to 2013.09.17 & \cite{willson03} & 0.83 & 0.70 & 0.34 \\
IRMB composite (dated 2016.11.21) & 1981.07.02 to 2016.10.31 & \cite{dewitte16} & 0.90 & 0.81 & 0.27 \\
PMOD composite (42\_65\_1608) & 1978.11.17 to 2016.08.02 & \cite{frohlich00,frohlich06} & 0.93 & 0.86 & 0.22 \\
\hline          
\end{tabular}
\end{sidewaystable}

\begin{figure}
\includegraphics[width=\textwidth]{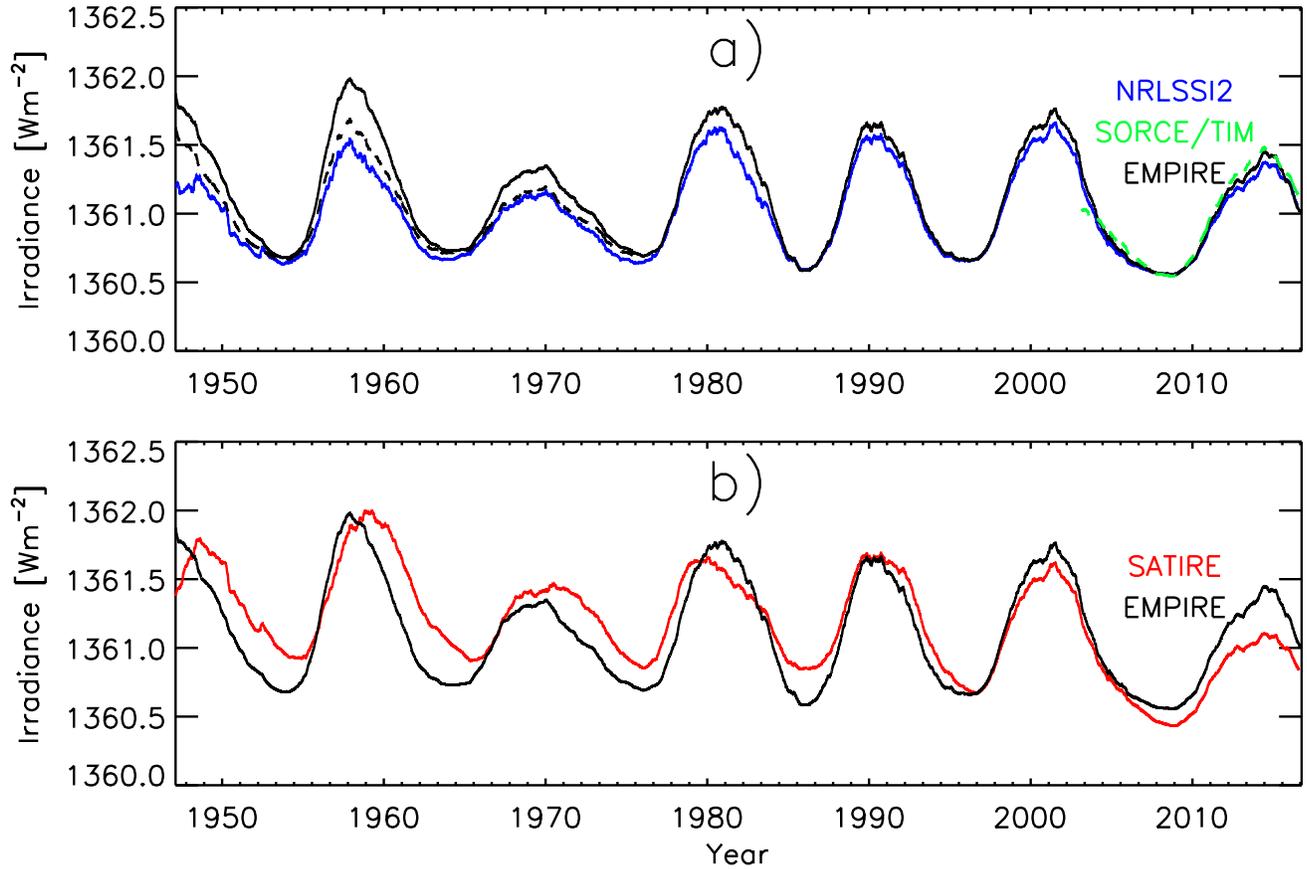}
\caption{The EMPIRE reconstruction of TSI over the full period of the model (black), compared to a) NRLSSI2 (blue) and TIM (green, dashed for visibility), and b) the composite of SATIRE-S and SATIRE-T (red). The 731-day moving average is illustrated. The NRLSSI2 and TIM time series are normalized to the EMPIRE time series at the 2008 solar cycle minimum and SATIRE to EMPIRE at the 1996 minimum. The black dashed line demonstrates the effect on EMPIRE if we generate the PSI composite as done in NRLSSI2 instead of using the \cite{balmaceda09} PSI composite. See Sect. \ref{tsicomparemodels} for details.}
\label{empssilongterm}
\end{figure}

\begin{figure}
\includegraphics[width=\textwidth]{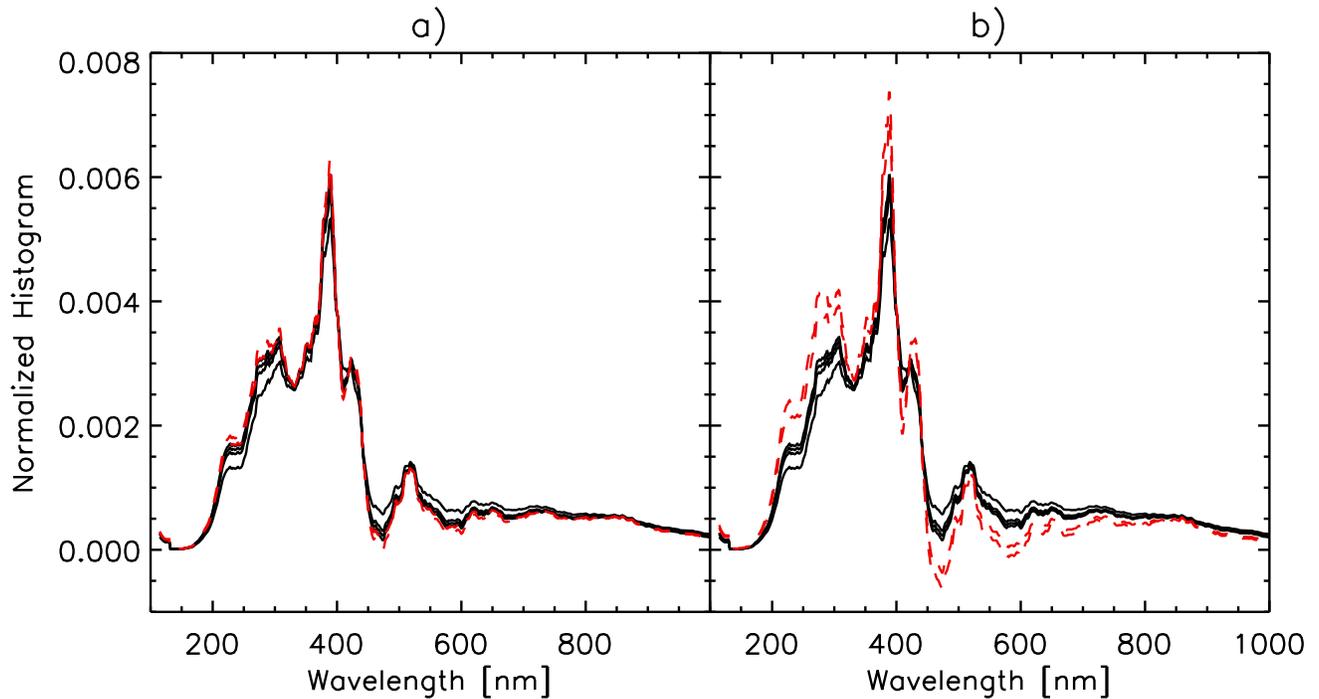}
\caption{a) For each solar cycle in the EMPIRE reconstruction, the spectral distribution of the change in solar irradiance between the minimum at the beginning of the cycle and the maximum. The 20 nm moving average over 115 to 1000 nm is illustrated. The red dashed lines correspond to solar cycles 19 and 20 and the black lines to solar cycles 21 to 24. b) The same, if we generate the PSI composite as done in NRLSSI2 instead of using the \cite{balmaceda09} PSI composite. See Sect. \ref{tsicomparemodels} for details.}
\label{sunspotareacalibrationfactor}
\end{figure}

\end{document}